\documentclass[10pt]{iopart}

\expandafter\let\csname equation*\endcsname\relax
\expandafter\let\csname endequation*\endcsname\relax

\usepackage{iopams,times}
\usepackage{graphicx,color,xcolor} 
\usepackage{amsmath}
\usepackage{tabularx}
\usepackage{braket}
\usepackage{bbold}
\usepackage{comment}
\usepackage{tikz}
\usepackage{subcaption}
\usepackage{hyperref}

\begin{document}

\title[Separate characterization of SPAM errors using single-qubit operations]{Separate and efficient characterization of state-preparation and measurement errors using single-qubit operations}

\author{Muhammad Qasim Khan$^1$, 
Leigh M. Norris$^2$\footnote{Present address: Quantinuum, Broomfield, Colorado, USA}, 
Lorenza Viola$^1$}
\address{$^1$ Department of Physics and Astronomy, Dartmouth College, Hanover, New Hampshire 03755, USA}
\address{$^2$ Johns Hopkins Applied Physics Laboratory, 11100 Johns Hopkins Road, Laurel, Maryland 20723, USA}
\ead{Lorenza.Viola@Dartmouth.edu}

\begin{abstract}
In many platforms, errors from state-preparation and measurement (SPAM) dominate {over} single-qubit gate errors. To inform further hardware improvements and the development of more effective SPAM mitigation strategies, it is necessary to separately characterize the error contributions from state-preparation (SP) and measurement (M). Here, we show how to construct a protocol that can efficiently and separately characterize the SP and M error parameters by using only high-fidelity single-qubit gates and repeated single-qubit measurements without reset. The measurements are assumed to be non-destructive and M errors are taken to be (spatially and temporally) uncorrelated and classical. Notably, the circuit depth of the protocol is independent of system size, and the target parameters may be characterized to a precision that is only limited by the number of experimental repetitions. We employ our protocol for the parallel characterization of SPAM errors on multiple qubits in the IBM Quantum Platform devices, where we find SP infidelities up to $6.57\%$ and readout assignment errors up to $19.1\%$. Using numerical simulations, we also demonstrate how measurement-error mitigation that does not properly account for SP errors generally leads to a biased estimate of measured observable expectation values.
\end{abstract}

\section{Introduction}

High-fidelity fiducial state preparation and measurement are central to any useful quantum science and technology application. Quantum-enhanced sensing protocols, for example, rely on specific state preparations and measurements to maximize the precision to which a target signal or parameter may be estimated \cite{Degen2017, Felix2018, riberi2023, riberi2025}, with the achievable performance being degraded in the presence of imperfect states and detectors \cite{Datta, Len2022}. SPAM errors have likewise been shown to significantly impact the performance of spectral estimation protocols for general spatio-temporally correlated noise based on quantum noise spectroscopy \cite{Szankowski_2017,PhysRevA.95.022121,PazSilvaFrames}, requiring appropriate protocol modification \cite{PhysRevApplied.22.024074}. Continued progress in both fundamental physics problems, including measurement-induced phase transitions in continuously monitored quantum many-body systems \cite{Skinner2019, Garratt2023, Murciano2023, PhysRevX.10.041020}, and computational applications ranging from quantum phase estimation \cite{Blunt2023} and two-qubit quantum teleportation \cite{IBM2025}, to syndrome extraction in quantum error correction {routines} \cite{Livingston2022, Acharya2025}, also crucially depend on accurate characterization of the classical readout from the system. The rate of SPAM errors varies significantly across different experimental platforms. While it is already remarkably small in some trapped-ion systems \cite{Quantinuum2022}, and hardware continues to improve, SPAM errors remain significant in superconducting transmon {devices} \cite{Laflamme2021, Klimov} and quantum-dot systems \cite{Blumoff2022, Takeda2024}, and also play an important role in {analog} Rydberg-atom simulators \cite{Endres2023,Bryce2024}.

A significant push in tackling SPAM errors has been made in the context of quantum characterization, verification, and validation methods \cite{Sandia}. In particular, SPAM-robust protocols for circuit-level noise characterization and gate calibration have been developed \cite{CompressedGST, NonMarkovianGST, White2020, hurant2024, FastBT}, as well as probabilistic methods for error mitigation \cite{QEM, Gupta2024,vandenBerg2023}, including recent extensions to drift-resilient schemes \cite{Uzdin}. These protocols are designed to be insensitive to SPAM errors, however, they do not necessarily separate the distinct contributions of SP and M errors. Under the assumption that SP errors may be taken to be negligible, techniques that are specifically designed to characterize and mitigate M errors for a {\em classical}, possibly correlated error model have been put forward \cite{PhysRevA.103.042605, PRXQuantum.2.040326, Ivashkov2024}, {along with approaches based on quantum detector tomography that can account for beyond-classical readout models \cite{PhysRevA.100.052315, DiGiovanni}. If both SP and M errors are assumed to contribute, recent work has shown that separate characterization is indeed possible;} yet, the protocols for achieving this separate characterization suffer from significant shortcomings in terms of both their required resources and scalability. In particular, for existing protocols that make use of two-qubit gates \cite{Laflamme2021,Yu2025}, the overall accuracy is limited to that of the two-qubit operation; furthermore, ancillary qubits are involved, 
making efficient parallelized characterization of multiple qubits difficult. Other protocols that utilize only single-qubit operations and self-consistent Bayesian estimation \cite{Mattias} rely on circuits that scale (polynomially) with the system size \textemdash thereby allowing for environmental noise to corrupt the estimates.

In this work, we present a {\em Quantum SPAM (QSPAM)} protocol which, provided that repeated measurements may be effected, is capable of separate, parallel characterization of single-qubit SP and M error parameters using circuits that do not scale with the system size. In its current form, the protocol is designed under the assumption of high-fidelity single-qubit gates and uncorrelated, classical M errors, {described} in terms of positive operator-valued measure (POVM) elements that are diagonal in the computational basis. With only eight parallelizable experiments per qubit, which consist of at most one single-qubit gate and two consecutive measurements, it is possible to characterize the initial SP errors and the readout POVMs for each qubit in the system, as well as the post-measurement state. A {\em simplified QSPAM (sQSPAM)} protocol comprising only five experiments per qubit is shown to suffice if the measurement (Kraus) operators, in addition to the POVM elements, are themselves assumed to be diagonal. 

The content is organized as follows. In Section \ref{sec::background}, we introduce the relevant assumptions and SPAM-error parameterizations, and develop the theory for the proposed sQSPAM and QSPAM protocols. In Section \ref{sec::verification}, the QSPAM protocol is implemented on the IBM Quantum Platform using Qiskit \cite{qiskit2024} to demonstrate that it is, in fact, capable of separately characterizing the SP and M parameters with a precision that is only limited by the available number of experimental shots. The QSPAM protocol is also used to characterize a large number of qubits in parallel, to show that these errors are significant in some of the qubits and, interestingly, the M operators are {\em not} diagonal for all qubit readouts. In Section \ref{sec::mit}, we present a simple SP-error-mitigation protocol that can correct for coherent SP errors, and revisit the standard M-error-mitigation protocol that is commonly employed on cloud-based platforms within the theoretical framework we employ in this work. In Section \ref{sec::GHZ}, we use numerical simulations of real devices to argue that {\em not} accounting for SP errors can generally lead to a {\em biased estimate} of observable expectation values of interest \textemdash possibly to an overestimate that may even be non-physical. We show this by preparing a many-qubit GHZ state and measuring a collective observable. Similar results from an actual IBM Q device are also presented in support of our claims. Section \ref{sec::discussion} concludes with a discussion of our main results and possible future research avenues. Several appendices are also included, providing additional derivations and technical detail.

\section{Theory background: SPAM characterization}
\label{sec::background}

To bypass the limitations of current SPAM characterization protocols, we use single-qubit gates that are {\em fast and high-fidelity}, and can be calibrated to be robust against SPAM errors \cite{hurant2024}, up to fidelities that are orders of magnitude higher than M errors on existing devices. Using only single-qubit operations and single-qubit readouts, we can construct a QSPAM  protocol that can estimate the POVM elements for a single qubit and uncorrelated multiqubit measurements, with circuits that are independent of the system size. Our key idea is to repeatedly and conditionally measure the same qubit without reset, which lets us disambiguate the SP and M error contributions, as we are going to explain next. 

\subsection{Standard SPAM characterization protocol}
\label{sec:std}

Let us first review the standard SPAM characterization protocol \cite{PhysRevA.103.042605}, for a single qubit (see Sec.\,\ref{sub::mmit} for the extension to a multiqubit setting). 
The qubit is {initially} prepared in the ground state, and the probability of finding it in the excited state, say ${\mathbb P}_{z_+\rightarrow z_-}$, is evaluated. Subsequently, the qubit is prepared in the excited state and the probability of finding it in the ground state, $\mathbb{P}_{z_-\rightarrow z_+}$, is evaluated. With respect to the computational basis, the resulting M errors are then expressed through {\em diagonal} POVM operators of the form 
\begin{align}
    \Pi_{z_+}\equiv \begin{bmatrix}
        \frac{1+\alpha_\text{M}+\delta}{2} & 0\\
        0 & \frac{1-\alpha_\text{M}+\delta}{2}
    \end{bmatrix},\qquad 
    \Pi_{z_-}\equiv \begin{bmatrix}
        \frac{1-\alpha_\text{M}-\delta}{2} & 0\\
        0 & \frac{1+\alpha_\text{M}-\delta}{2}
    \end{bmatrix},
    \label{eq::povm1}
\end{align}
which satisfy the completeness condition $\Pi_{z_+}+\Pi_{z_-}=\mathbb{1}$. Here, the parameter $\alpha_\text{M}$ represents the fidelity of the readout, whereas $\delta$ represents the asymmetry in the fidelity of readout between the ground state and the excited state \cite{Yu2025,PhysRevApplied.22.024074,PhysRevA.103.042605,Zoltan}. The positivity requirement of these operators mandates the following constraints on the parameters: 
\begin{align}
\label{eq::measbounds}
    0\leq\alpha_\text{M}\leq1 ,\qquad 
    \alpha_\text{M}-1\leq\delta\leq1-\alpha_\text{M}. 
\end{align}
The above POVM elements represent {\em classical readout errors}, because they are taken to be diagonal and, as such, they represent readout assignment errors on the classical readout registers. Although nonclassical coherent errors -- which render the POVM non-diagonal -- may {\em a priori} be present, we assume they are typically negligible compared to the classical contribution. This is consistent with existing results \cite{Zoltan, DiGiovanni} and aligns with our assumption of high-fidelity single-qubit gates, as such errors are largely attributable to imperfect basis rotations. A quantitative analysis of the departure from a diagonal POVM model can be found in the cited literature. Furthermore, if the qubit is part of a multi-qubit device, by considering only single-qubit POVMs and constant error parameters, we implicitly assume that the readouts are spatially and temporally uncorrelated across different qubits.

Let us assume that preparation of the intended $+1$ eigenstate of $\sigma_z$ (which we will denote $|z_+\rangle$) results in a faulty initial state given by 
\begin{align}
\label{eq::1qrho0}
    \rho=\frac{1}{2}\bigg(\mathbb{1}+\!\sum_{i=x,y,z}\alpha_\text{SP}^i\sigma_i\bigg),
\end{align}
in terms of SP parameters:
\begin{align}
\label{eq::SPbounds}
    \alpha^z_\text{SP}\in(0,1], && \alpha_\text{SP}^{x},\alpha_\text{SP}^{y}\in(-1,1), && \sum_{i=x,y,z}(\alpha^i_\text{SP})^2\leq 1.
\end{align}
Thus, we may parameterize the faulty SP with a vector $\vec{\alpha}_\text{SP}\equiv (\alpha_\text{SP}^x$, $\alpha_\text{SP}^y$, $\alpha_\text{SP}^z)$, with $\vec{\alpha}_\text{SP}=(1,0,0)$ corresponding to the ideal $|z_+\rangle$ state.

With the POVM elements in Eq.\,\eqref{eq::povm1} representing the faulty measurements, the standard SPAM characterization protocol uses the probabilities of preparing the qubit in the ground state and measuring it in the excited state (and vice-versa); that is, 
\begin{align}
\label{eq::p0to1}
    \mathbb{P}_{z_+\rightarrow z_-}&=\frac{1}{2}\left(1-\alpha_\text{M}\alpha_\text{SP}^z-\delta\right), \\
\label{eq::p1to0}
    \mathbb{P}_{z_-\rightarrow z_+}&=\frac{1}{2}\left(1-\alpha_\text{M}\alpha_\text{SP}^z+\delta\right).
\end{align}
These two equations allow us to solve for $\delta$ and $\alpha_\text{M}\alpha_\text{SP}^z$ as a {\em product}. Typically, the next step is to assume that the SP is sufficiently accurate  such that $\alpha^z_\text{SP}\approx 1$. By then letting $\alpha_\text{M}\alpha_\text{SP}^z\approx \alpha_\text{M}$, a {\em confusion matrix} $A$ for the readout is constructed,
\begin{align}
\label{eq::confused}
    A&=\begin{bmatrix}
        \frac{1+\alpha_\text{M}+\delta}{2} & \frac{1-\alpha_\text{M}+\delta}{2}\\
        \\
        \frac{1-\alpha_\text{M}-\delta}{2} & \frac{1+\alpha_\text{M}-\delta}{2}
    \end{bmatrix}\approx \begin{bmatrix}
        \mathbb{P}_{z_+\rightarrow z_+} & \mathbb{P}_{z_-\rightarrow z_+}\\
        \\
        \mathbb{P}_{z_+\rightarrow z_-} & \mathbb{P}_{z_-\rightarrow z_-}
    \end{bmatrix},
\end{align}
which may be inverted to obtain estimates for the measured observable expectation values. However, as we will demonstrate, 
$\alpha^z_\text{SP}$ can significantly differ from its ideal unit value in reality -- which necessitates additional single-qubit experiments.

\subsection{sQSPAM protocol}
\label{sub:diagonal}

To separately characterize the SPAM-errors $\alpha_\text{M}$, $\delta$, $\alpha^x_\text{SP}$, $\alpha^y_\text{SP}$, $\alpha^z_\text{SP}$, we propose to carry out an additional single-qubit experiment, in which the qubit is measured {\em twice without reset}. We model the measurement as \emph{efficient}, meaning that the post-measurement state is pure if the pre-measurement state is pure. If so, the measurement map that updates the state of the qubit conditioned on each outcome $z_\pm$ is described by a \emph{single} Kraus operator $M_{z_\pm}$, rather than a sum over multiple Kraus operators as in the general case. The operators satisfy $M_{z_\pm}^\dagger M_{z_\pm}= \Pi_{z_\pm}$, corresponding to the POVM in Eq.\,\eqref{eq::povm1}, and $M_{z_+}^\dagger M_{z_+} + M_{z_-}^\dagger M_{z_-} = \mathbb{1}$. Explicitly, in the computational basis, we may write 
\begin{align}
\label{eq::Mels}
    M_{z_\pm}&\equiv 
    \begin{bmatrix}
        M^\pm_{z_+z_+} & M^\pm_{z_+z_-} \\
        M^\pm_{z_-z_+} & M^\pm_{z_-z_-}
    \end{bmatrix}.
\end{align}
Crucially, we place {\em no diagonal constraint} on the measurement operators themselves. This allows our formalism to capture non-unitary back-action effects inherent to dispersive readout, such as measurement-induced dephasing or bit-flip events via the off-diagonal elements of $M_{z_\pm}$. Consequently, the protocol does not assume ideal state preservation; rather, it utilizes the effective post-measurement state derived from the efficient measurement operators.

While this set of Kraus operators (referred to subsequently as just M operators) is highly {non-unique}, some constraints arise from the condition that the resulting POVM elements be diagonal. Specifically,  this requirement leads to one of the coefficients being fixed, $M^\pm_{z_+z_-}=-M^\pm_{z_-z_-}\tfrac{M^{\pm*}_{z_-z_+}}{M^{\pm*}_{z_+z_+}}$, with the resulting POVM elements taking the form 
\begin{align}
\label{eq::POVMgeneralM}
    \Pi_{z_+}&=
    \begin{bmatrix}
        |M^+_{z_+z_+}|^2\left(1+\tfrac{|M^+_{z_-z_+}|^2}{|M^+_{z_+z_+}|^2}\right) & 0\\
        0 & |M^+_{z_-z_-}|^2\left(1+\tfrac{|M^+_{z_-z_+}|^2}{|M^+_{z_+z_+}|^2}\right)
    \end{bmatrix},
\end{align}
and  $\Pi_{z_-}=\mathbb{1}-\Pi_{z_+}$. In turn, the positive semi-definite nature of POVM elements enforces the following constraints on the elements of the M operators:
\begin{align}
\label{eq::POVMbound}
     0 \leq |M^+_{z_-z_-}|^2 < |M^+_{z_+z_+}|^2\leq 1, && 0 \leq |M^+_{z_-z_+}|^2\leq 1 - |M^+_{z_+z_+}|^2.
\end{align}
As a result, in the most general scenario for a diagonal POVM measurement on a single qubit, an extra complex parameter $M^+_{z_-z_+}$ arises as compared to Eq.\,\eqref{eq::povm1}, although reconstructing the POVM elements still requires estimating {\em two} real parameters (namely, the diagonal components of the POVM). Explicitly, we can rewrite the parametrization for each of the M operators in Eq.\,\eqref{eq::Mels} in terms of three amplitudes and two unique phases 
for each of the M. Taking into account the normalization constraint, estimating the two M operators requires an estimation of {\em eight} independent parameters:
\begin{align}
    M_{z_\pm}&=
    \begin{bmatrix}
        |M^\pm_{z_+z_+}|e^{i\phi^{(\pm)}_{z_+z_+}} & -|M^\pm_{z_-z_-}|\tfrac{|M^{\pm}_{z_-z_+}|}{|M^{\pm}_{z_+z_+}|}e^{i(\phi^{(\pm)}_{z_+z_+}+\phi^{(\pm)}_{z_-z_-})} \\
        |M^\pm_{z_-z_+}| & |M^\pm_{z_-z_-}|e^{i\phi^{(\pm)}_{z_-z_-}}
    \end{bmatrix}.
    \label{eq::Mgen}
\end{align}

Suppose, however, that we can work under the simplifying assumption that the M operators, not only the POVM elements, are themselves {\em diagonal}: that is, we let $M^+_{z_-z_+}=0$. In such a case, we may use a simpler notation $M_{z_\pm} \mapsto M_\pm$ and, up to an irrelevant global phase, we may re-parametrize the operators in Eq.\,\eqref{eq::Mgen} in terms of just {\em four} real parameters:  
\begin{align}
    M_+=\begin{bmatrix}
        \sqrt{\frac{1+\alpha_\text{M}+\delta}{2}} & 0\\
        0 & e^{i\phi_+}\sqrt{\frac{1-\alpha_\text{M}+\delta}{2}}
    \end{bmatrix},\quad 
    M_-=\begin{bmatrix}
        \sqrt{\frac{1-\alpha_\text{M}-\delta}{2}} & 0\\
        0 & e^{i\phi_-}\sqrt{\frac{1+\alpha_\text{M}-\delta}{2}}
    \end{bmatrix} .
    \label{eq::diagM}
\end{align}
The post-measurement state $\rho^{(+)}_\text{PM}$, conditional on the outcome $z_+$, is then given by  
\begin{align}
\label{eq::rhoPM}
    \rho^{(+)}_\text{PM}&=
    \frac{M_+\rho M_+^\dagger}{\text{Tr}\left[\Pi_{z_+}\rho\right]},\\
    \label{eq::pmstate+}
    \rho^{(+)}_\text{PM}&=
    \begin{bmatrix}
        \frac{(1+\alpha_\text{SP}^z)(1+\alpha_\text{M}+\delta)}{2(1+\alpha_\text{M}\alpha_\text{SP}^z+\delta)} & \frac{e^{-i\phi_+}(\alpha_\text{SP}^x-i\alpha_\text{SP}^y)\sqrt{(1+\delta)^2-\alpha_\text{M}^2}}{2(1+\alpha_\text{M}\alpha_\text{SP}^z+\delta)}\\
        \frac{e^{i\phi_+}(\alpha_\text{SP}^x+i\alpha_\text{SP}^y)\sqrt{(1+\delta)^2-\alpha_\text{M}^2}}{2(1+\alpha_\text{M}\alpha_\text{SP}^z+\delta)}& \frac{(1-\alpha_\text{SP}^z)(1-\alpha_\text{M}+\delta)}{2(1+\alpha_\text{M}\alpha_\text{SP}^z+\delta)}
    \end{bmatrix}.
\end{align}
Therefore, the probability of the qubit remaining in the ground state after two consecutive measurements is
\begin{align}
    \mathbb{P}_{z_+\rightarrow z_+\rightarrow z_+}&=\text{Tr}\Bigg[\Pi_{z_+}\frac{M_+\rho M_+^\dagger}{\text{Tr}\left[\Pi_{z_+}\rho\right]}\Bigg],
\end{align}
and the expression can be evaluated in terms of the parameters under consideration as
\begin{align}
\label{eq::p0to0to0}
    \mathbb{P}_{z_+\rightarrow z_+\rightarrow z_+}&=\frac{\alpha_\text{M}^2+2\alpha_\text{M}\alpha_\text{SP}^z(1+\delta)+(1+\delta)^2}{2(1+\alpha_\text{M}\alpha_\text{SP}^z+\delta)}.
\end{align}
Taken together, Eqs.\,\eqref{eq::p0to1}, \eqref{eq::p1to0} and \eqref{eq::p0to0to0} are sufficient to characterize the parameters $\alpha_\text{M}$, $\delta$ and $\alpha_\text{SP}^z$ independently. 

Importantly, the remaining SP parameters $\alpha_\text{SP}^x$ and $\alpha_\text{SP}^y$ can also be characterized by simply repeating the experiments of the form in Eq.\,\eqref{eq::p0to1} with the qubit prepared in the $+1$ eigenstate of $\sigma_x$ and $\sigma_y$, and still measured in the $z$ eigenbasis. The preparation is made using the fiducial state $\rho$ provided by the system [Eq.\,\eqref{eq::1qrho0}], and applying a change of basis operation using the Hadamard-gate $H$ and the $\sqrt{X}$-gate, also called the $SX$ gate \cite{qiskit2024}, to obtain the readout probabilities
\begin{align*}
    \mathbb{P}_{x_+\rightarrow z_-}=\text{Tr}\left[\Pi_{z_-}H\rho H\right],\qquad 
    \mathbb{P}_{y_+\rightarrow z_-}=\text{Tr}\left[\Pi_{z_-}SX^\dagger\rho SX\right],
\end{align*}
respectively. The full set of measured probabilities obtained in this way defines a {\em simplified QSPAM} (sQSPAM) protocol, which affords a separate characterization of SP and M errors under the assumption of diagonal M operators:
\begin{align}
    \mathbb{P}_{z_+\rightarrow z_-}&=\frac{1}{2}\left(1-\alpha_\text{M}\alpha_\text{SP}^z-\delta\right),\\
    \mathbb{P}_{z_-\rightarrow z_+}&=\frac{1}{2}\left(1-\alpha_\text{M}\alpha_\text{SP}^z+\delta\right),\\
    \label{eq::sqspamx}
    \mathbb{P}_{x_+\rightarrow z_-}&=\frac{1}{2}\left(1-\alpha_\text{M}\alpha_\text{SP}^x-\delta\right),\\
        \label{eq::sqspamy}
    \mathbb{P}_{y_+\rightarrow z_-}&=\frac{1}{2}\left(1-\alpha_\text{M}\alpha_\text{SP}^y-\delta\right),\\
    \mathbb{P}_{z_+\rightarrow z_+\rightarrow z_+}&=\frac{\alpha_\text{M}^2+2\alpha_\text{M}\alpha_\text{SP}^z(1+\delta)+(1+\delta)^2}{2(1+\alpha_\text{M}\alpha_\text{SP}^z+\delta)}.
\end{align}
In the parameter bounds defined in Eqs.\,\eqref{eq::measbounds} and \eqref{eq::SPbounds}, there exists a unique solution to the above set of equations for the parameters $\{\alpha_\text{M},\delta,\alpha_\text{SP}^x,\alpha_\text{SP}^y,\alpha_\text{SP}^z\}$.

Estimating the phases $\phi_+$ and $\phi_-$ is not necessary to reconstruct the POVMs. These phases may be estimated through measurements of the coherence elements of the post-measurement states $\rho^{(+)}_\text{PM}$ and $\rho^{(-)}_\text{PM}$, if the M operators are to be estimated.

\subsection{QSPAM protocol with general efficient measurements}
\label{sub::general} 

Assuming that the M operators are diagonal is a simplification which, {\em a priori}, cannot be justified. It is thus both desirable and important to determine whether estimating all the desired SPAM parameters, $\{\alpha_\text{M}, \delta, \alpha_\text{SP}^x, \alpha_\text{SP}^y, \alpha_\text{SP}^z\}$, remains possible by assuming the most general form of M operators compatible with diagonal POVM elements, given in Eq.\,\eqref{eq::Mgen}. Let us introduce a parameter that quantifies the deviation of M operators from diagonal form:
\begin{equation}
 \epsilon \equiv |M^+_{z_-z_+}|^2/ |M^+_{z_+z_+}|^2.
\label{eq::eps}
\end{equation}
The M parameters can then be related to the most general form of the POVM by comparing Eqs.\,\eqref{eq::povm1} and \eqref{eq::POVMgeneralM}, which yields:
\begin{align}
\label{eq::aMgeneral}
    \alpha_\text{M}&\equiv\left(|M_{z_+z_+}^+|^2-|M_{z_-z_-}^+|^2\right)\left(1+\epsilon\right),\\
\label{eq::deltageneral}
    \delta&\equiv\left(|M_{z_+z_+}^+|^2+|M_{z_-z_-}^+|^2\right)\left(1+\epsilon\right)-1.
\end{align}
In this case, the post-measurement state after the first measurement, conditioned on a $z_+$ outcome, follows from $\rho_\text{PM}^{(+)}$ given in Eq.\,\eqref{eq::rhoPM}, with $M_+$ replaced by the (non-diagonal) operator $M_{z_+}$. Accordingly, the probability $\mathbb{P}_{z_+\rightarrow z_+ \rightarrow z_+}$ is found to be 
\begin{multline}
\label{eq::p000}
    \mathbb{P}_{z_+ \rightarrow z_+ \rightarrow z_+} = \frac{1}{2(1+\epsilon)(1+\alpha_{\rm M}\alpha_{\rm SP}^z+\delta)}\bigg[\alpha_{\rm M}^2(1-\epsilon) + 2\alpha_{\rm M}\alpha_{\rm SP}^z(1+\delta) + (1+\delta)^2(1+\epsilon) \\
    - 2\alpha_{\rm M}\sqrt{\epsilon\big((1+\delta)^2-\alpha_{\rm M}^2\big)}\,\Big(\alpha_{\rm SP}^x\cos\phi^{(+)}_{z_-z_-} - \alpha_{\rm SP}^y\sin\phi^{(+)}_{z_-z_-}\Big)\bigg].
\end{multline}
Note that the above expression, unlike Eq.\,\eqref{eq::p0to0to0}, generally depends on the phase $\phi^{(+)}_{z_-z_-}$, unless $\epsilon =0$. We can remove this phase-dependence through $z$-rotation operations on the fiducial state of the qubit parameterized by the angle $\theta$, $R_z(\theta)$. Specifically, we introduce three additional experiments and, altogether, measure the following set of probabilities:
\begin{multline*}
    \Big\{\mathbb{P}_{z_+\rightarrow z_-}, \mathbb{P}_{z_-\rightarrow z_+}, \mathbb{P}_{x_+\rightarrow z_-}, \mathbb{P}_{y_+\rightarrow z_-}, \\
    \mathbb{P}_{z_+\rightarrow z_+\rightarrow z_+}^{\theta=0}, \mathbb{P}_{z_+\rightarrow z_+\rightarrow z_+}^{\theta=\pi}, \mathbb{P}_{z_-\rightarrow z_+\rightarrow z_+}^{\theta=0}, \mathbb{P}_{z_-\rightarrow z_+\rightarrow z_+}^{\theta=\pi}\Big\}.
\end{multline*}
Altogether, the resulting set of experiments constitutes our {\em QSPAM protocol}. 
\begin{figure}[t]
    \centering
    \includegraphics[width=.55\linewidth]{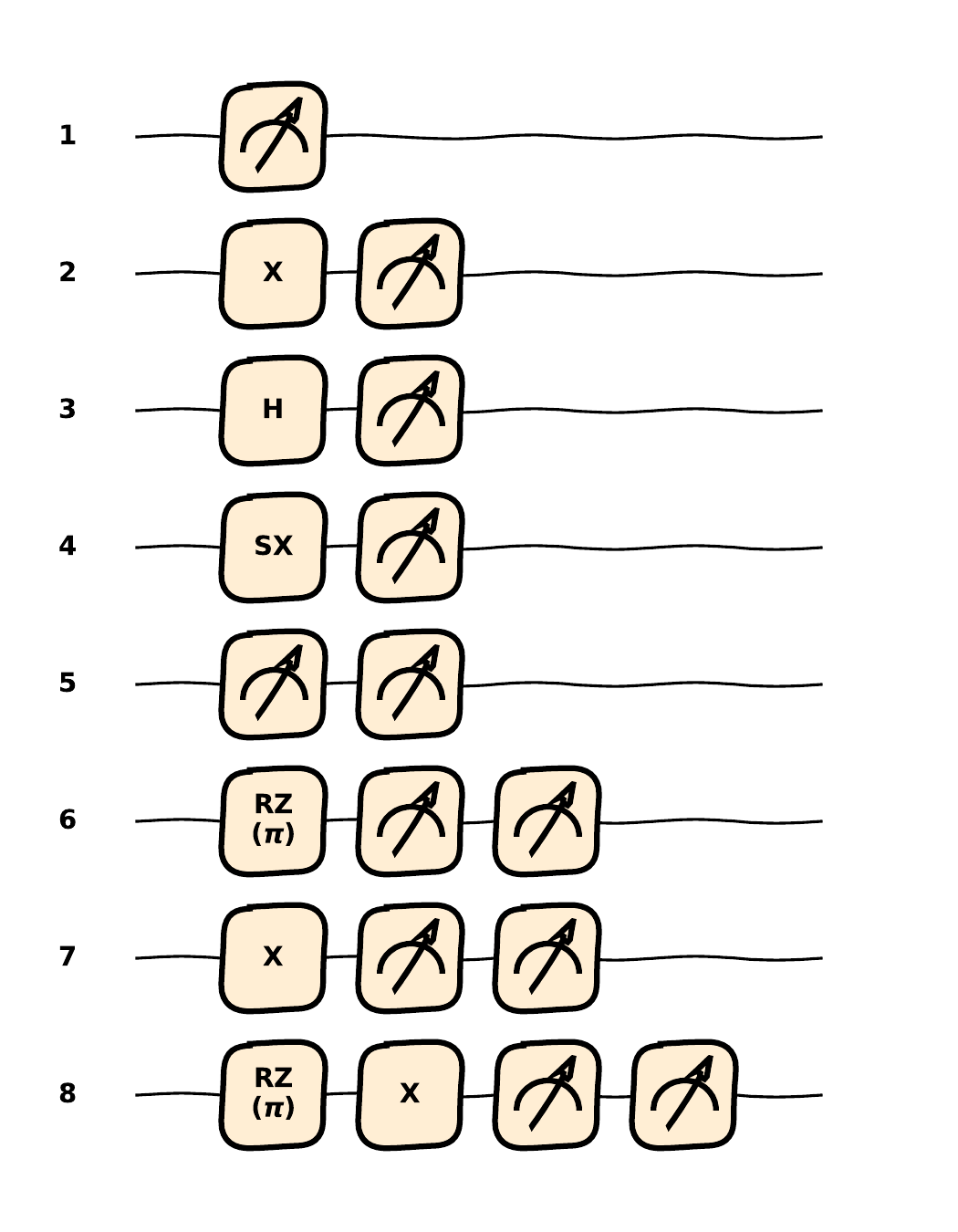}
    \vspace*{-4mm}
    \caption{{\small
    Pictorial representation of the {\em eight} experiments required for the separate characterization of SP and M parameters by using the proposed QSPAM protocol.}}
    \label{fig::circs}
\end{figure}

While the QSPAM quantum circuits are displayed in Fig.\,\ref{fig::circs}, all the relevant probabilities may be obtained from the following explicit expressions:
\begin{align}
\label{eq::p0to1F}
    \mathbb{P}_{z_+\rightarrow z_-}&=\frac{1}{2}\left(1-\alpha_\text{M}\alpha_\text{SP}^z-\delta\right), \qquad 
    \mathbb{P}_{z_-\rightarrow z_+}=\frac{1}{2}\left(1-\alpha_\text{M}\alpha_\text{SP}^z+\delta\right), \\
        \mathbb{P}_{x_+\rightarrow z_-}&=\frac{1}{2}\left(1-\alpha_\text{M}\alpha_\text{SP}^x-\delta\right), \qquad  
    \mathbb{P}_{y_+\rightarrow z_-}=\frac{1}{2}\left(1-\alpha_\text{M}\alpha_\text{SP}^y-\delta\right),
\end{align}
along with the following combinations:
\begin{align}    
 \!\!\!\!\!  \mathbb{P}^{\theta_z=0}_{z_+\rightarrow z_+\rightarrow z_+}+\mathbb{P}^{\theta_z=\pi}_{z_+\rightarrow z_+\rightarrow z_+}&=\frac{\alpha_\text{M}^2(1-\epsilon)+2\alpha_\text{M}\alpha_\text{SP}^z(1+\delta)+(1+\delta)^2(1+\epsilon)}{(1+\alpha_\text{M}\alpha_\text{SP}^z+\delta)(1+\epsilon)},
\end{align}

\begin{align}   
\label{eq::p1to0to0piF}
 \!\!\!\!\!  \mathbb{P}^{\theta_z=0}_{z_-\rightarrow z_+\rightarrow z_+}+\mathbb{P}^{\theta_z=\pi}_{z_-\rightarrow z_+\rightarrow z_+}&=\frac{\alpha_\text{M}^2(1-\epsilon)-2\alpha_\text{M}\alpha_\text{SP}^z(1+\delta)+(1+\delta)^2(1+\epsilon)}{(1-\alpha_\text{M}\alpha_\text{SP}^z+\delta)(1+\epsilon)}.
\end{align}

The above set of nonlinear equations permits multiple solutions for the six parameters 
$$\{\alpha_\text{SP}^x,\alpha_\text{SP}^y,\alpha_\text{SP}^z,\alpha_M,\delta,\epsilon\},$$
with $\epsilon$ defined in Eq.\,\eqref{eq::eps}. It is thus necessary to choose the unique solution that resides within the physically permitted bounds on the M parameters, Eq.\,\eqref{eq::measbounds}, and the SP parameters, Eq.\,\eqref{eq::SPbounds}. The parameters can be solved for analytically by using the equations above, see \ref{app:analyticSPAM}.

In practice, the parameter estimation is carried out using a bounded and weighted numerical solver for Eqs.\,\eqref{eq::p0to1F}-\eqref{eq::p1to0to0piF}. The weights are set to be the reciprocals of the variances for each readout probability measured in the experiments. When a valid solution $\vec{x}_0$ is found for the SPAM parameters, the corresponding covariance matrix is approximated by the first-order correction to the estimated parameters, which is represented in terms of the residual Jacobian $J(\vec{x}_0)$. The covariance matrix is proportional to the reciprocal of the weight matrix $W^{-1}$, according to \cite{Weisberg2014}, 
 $   \textbf{Cov} (\vec{x}_0)=\left[J(\vec{x}_0)^TWJ(\vec{x}_0)\right]^{-1}$,
which is a diagonal matrix of weights. Importantly, since the variances of the estimated readout probabilities scale as $\sim \nu^{-1/2}$ in the number of shots $\nu$, it follows that the elements of the covariance matrix itself, that are linear combinations of these variances, also scale as $\sim \nu^{-1/2}$.

One may wonder to what extent our protocol may encounter a gauge freedom ambiguity, similar to gate set tomography (GST) \cite{Nielsen2021gatesettomography}. As also discussed in \cite{Laflamme2021}, the assumption of perfect gates we have made in constructing the protocol fixes the gauge to be the one that maps single-qubit gates to the perfect single-qubit gates. Care in addressing gauge freedom is needed in a scenario where the single-qubit gate infidelities dominate over the SPAM errors. In such a case, 
it is possible to modify the sQSPAM protocol in such a way that SP and M errors can still be separately characterized. A brief discussion is provided in \ref{app:FG}.

\section{Validation of the QSPAM protocol}
\label{sec::verification}

\subsection{Inferring state-preparation and measurement parameters} 
\label{sub::spNm}

As a first step, we report the verification of the sQSPAM protocol described in Sec.\,\ref{sub:diagonal} on a real device. To show that SP and M errors may be separately characterized, SP errors are artificially introduced in every characterization circuit of Fig.\,\ref{fig::circs},  by inserting a rotation $R_x(\phi)$ after the qubit initialization, and progressively increasing $\phi$. As a signature of the protocol working, within a short time span, we expect to see that the characterized $\alpha_\text{M}$ remains the same, whereas the SP error changes in correspondence with the injected error.

A simple calculation shows that an $R_x(\phi)$ rotation on the faulty initial ground state yields a rotated state $\overline{\rho}$ that may be expressed in terms of new SP parameters as follows:
\begin{align}
    \overline{\rho}&=R_x(\phi)\rho R_x^\dagger(\phi)=\frac{1}{2}\left[\mathbb{1}+\overline{\alpha}_\text{SP}^x(\phi)\sigma_x + \overline{\alpha}_\text{SP}^y(\phi)\sigma_y+\overline{\alpha}_\text{SP}^z(\phi)\sigma_z\right],\\
    \overline{\alpha}_\text{SP}^x(\phi) &\equiv \alpha_\text{SP}^x,\notag\\
    \overline{\alpha}_\text{SP}^y(\phi) &\equiv \alpha_\text{SP}^y\cos\phi-\alpha_\text{SP}^z\sin\phi,\notag\\
    \overline{\alpha}_\text{SP}^z(\phi) &\equiv \alpha_\text{SP}^z\cos\phi+\alpha_\text{SP}^y\sin\phi.\notag
\end{align}
Over a short period of time (as long as it takes to execute a circuit on our device), we expect the native parameters $\alpha_\text{SP}^{x},\,\alpha_\text{SP}^{y},\,\alpha_\text{SP}^{z}$ and $\alpha_\text{M}$ to remain stable, and a linear change in $\phi$ to induce a predominantly cosine shift in the characterized $\hat{\overline{\alpha}}_\text{SP}^z$, that we should be able to reconstruct using the QSPAM protocol. For this demonstration, it is sufficient to estimate $\overline{\alpha}_\text{SP}^z(\phi)$ independently from $\alpha_\text{M}$. For this reason, although possible in principle, we chose not to characterize the $\overline{\alpha}_\text{SP}^x(\phi)$ and $\overline{\alpha}_\text{SP}^y(\phi)$ parameters.

\begin{figure}[t]
    \centering
    \includegraphics[width=0.8\linewidth]{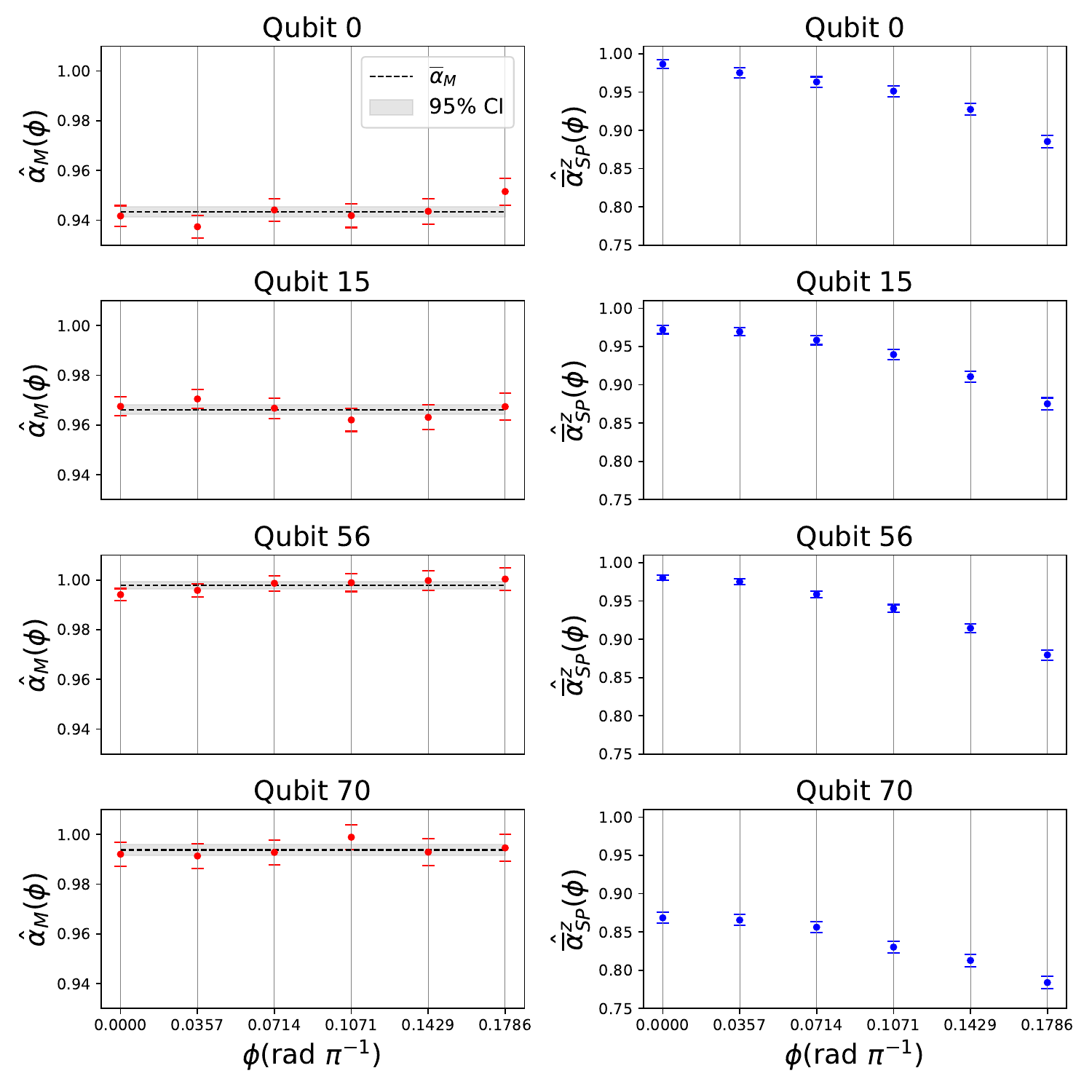}
    \vspace*{-2mm}
    \caption{{\small 
    Characterization of SP and M  parameters for arbitrarily selected qubits in the \texttt{ibm\_brisbane} device, where the characterization is implemented over $2^{14}$ shots, with no delay in between each execution of a circuit. The $x$-axis is the amount of injected SP error in radians, the left column shows the estimated M-error parameter $\hat{\alpha}_\text{M}$, and the right column is the SP component along $z$, $\hat{\overline{\alpha}}_\text{SP}^{z,(i)}$. The error bars on each data-point represent a $95\%$ confidence interval. The dashed line in the left column shows the average value for $\hat{\alpha}_\text{M}(\phi)$ as the injected SP error is increased, and the shaded area represents the $95\%$ confidence interval for the estimated mean value. The plot shows that the sQSPAM protocol successfully characterizes the two sources of error independently, as predicted.}}
    \label{fig:almalsp}
\end{figure}

The results of this characterization on IBM-Q's \texttt{ibm\_brisbane} device are presented in Fig.\,\ref{fig:almalsp}. In this demonstration, we injected an identical SP errors into qubits $i\in \{0,15,56, 70\}$ and applied the sQSPAM characterization protocol {\em in parallel} on these qubits to estimate the parameters $\{\hat{\alpha}^{(i)}_\text{M}$, $\hat{\delta}_{(i)}, \hat{\overline{\alpha}}_\text{SP}^{z,(i)}\}$ for all $i$. Consistent with the simplifying assumptions underlying sQSPAM, we took the M operators to be diagonal, as in Eq.\,\eqref{eq::diagM}. While the presence of non-negligible non-diagonal elements of the M operators would lower the accuracy of the characterized $\hat{\alpha}^{(i)}_\text{M}$ and $\hat{\delta}_i$ (see Eq.\,\eqref{eq::aMgeneral}), it would not change the stability of $\hat{\alpha}^{(i)}_\text{M}$ and the variation expected for $\hat{\alpha}_\text{SP}^{z,(i)}$ under changing SP error. The data in Fig.\,\ref{fig:almalsp} very clearly demonstrates that, despite some fluctuations, the M parameter remains {\em unchanged} (within a $95\%$ confidence interval of the mean value) even when the initial state is rotated by $\pi/5.6$ radians, whereas the estimates $\hat{\overline{\alpha}}^{z,(i)}_\text{SP}(\phi)$ follow a cosine, as we predicted.

\subsection{Characterizing measurement operators}
\label{sub::devicechar}

As a next important question, we wish to determine whether or not the measurement operators on a real quantum device may be assumed to be diagonal. To this end, we deploy the full QSPAM characterization protocol described in Sec.\,\ref{sub::general}. Quantitatively, the M operator may be taken to be diagonal if the correction factor arising from their off-diagonal matrix elements in Eqs.\,\eqref{eq::aMgeneral}- \eqref{eq::deltageneral} is sufficiently small.

\begin{figure}[t]
    \centering
    \includegraphics[width=\linewidth]{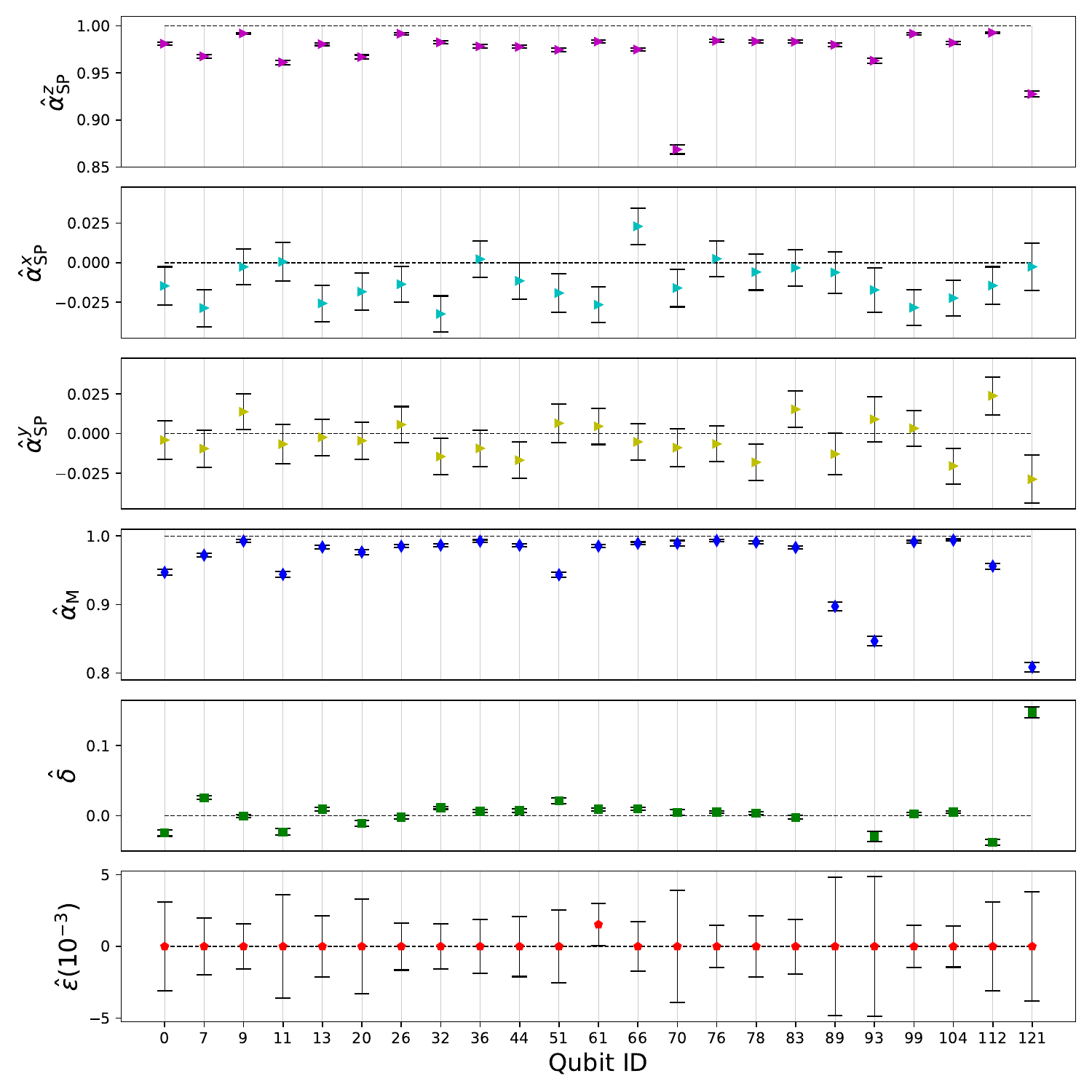}
    \vspace*{-5mm}
    \caption{{\small 
    Characterization of the SPAM parameters assuming non-diagonal M operators on the \texttt{ibm\_brisbane} device. The experiments were conducted on a set of qubits that are disconnected and spread across the device, with the parameters estimated over $2^{15}$ shots. The results demonstrate a level of SP errors in qubits across the 127-qubit device that is {\em not} negligible. The last row shows the contribution of the off-diagonal element of the M operator through the parameter $\epsilon$ from Eq.\,\ref{eq::eps}. For most of the qubits we characterized, $\hat{\epsilon}\approx0$; for Qubit $61$, however, we find $\hat{\epsilon}_{(61)}=0.0015\pm0.0007$, implying {\em non-diagonal} M operators.}}
    \label{fig:QSPAMbris}
\end{figure}

The results of a complete characterization obtained in the \texttt{ibm\_brisbane} device under the assumption of general M operators  are shown in Fig.\,\ref{fig:QSPAMbris}. The characterization shows the extent of variability of SPAM-errors across the device. For instance, Qubit-$121$ suffers from both significant SP and M errors with parameters $\hat{\delta}_{(121)}=0.1476\pm0.0037$, $\hat{\alpha}^{(121)}_\text{M}=0.8088\pm0.0033$ and $\hat{\alpha}^{z,(121)}_\text{SP}=0.9276\pm0.0016$. On the other hand, Qubit-$9$ is a high-quality qubit, with $\hat{\delta}_{(9)} = 0\pm0.0009$, $\hat{\alpha}^{(9)}_\text{M}=0.9928\pm0.0009$ and $\hat{\alpha}^{z,(9)}_\text{SP}=0.9919\pm0.0005$. For both of the aforementioned qubits, the correction parameter in Eq.\,\eqref{eq::eps} is effectively 0 ($\hat{\epsilon}_{(9)}=0\pm0.0008$ and $\hat{\epsilon}_{(121)}=0\pm0.0019$), hence the M operators may be considered diagonal. Within the characterized qubit set, Qubit-$61$ appears to be the only one with a significant off-diagonal component of the M operator, $\hat{\epsilon}_{(61)}=0.0015\pm0.0007$.

\section{SPAM mitigation}
\label{sec::mit}
\subsection{Mitigating state-preparation error}
\label{sub::spmit}

After estimating the Bloch vector of the faulty initial state preparation, $\hat{\vec{\alpha}}_\text{SP}\equiv ({\hat\alpha}_\text{SP}^x, \hat{\alpha}_\text{SP}^y, \hat{\alpha}_\text{SP}^z)$, it becomes possible to mitigate SP errors at the beginning of any experiment through an appropriate corrective rotation. Let the ideal ground state correspond to the Bloch vector $\vec{g} \equiv (0, 0, 1)$. Native operations on quantum computing platforms are often limited, and a rotation about an arbitrary axis given by $\hat{\vec{\alpha}}_\text{SP}\times\vec{g}/|\hat{\vec{\alpha}}_\text{SP}\times\vec{g}|$ has to be transpiled into these native operations. Specifically, we present an operation to correct for the SP errors on the IBM-Q devices, whereby $X$, $R_z(\phi)$ and $R_x\left(\frac{\pi}{2}\right)$ are the native single-qubit operations. By minimizing the use of gates other than the virtual $R_z(\phi)$, we can ensure a high-fidelity SP error mitigation. We claim that the rotation 
\begin{align}
R\big(\hat{\vec{\alpha}}_\text{SP}\big)& \equiv R_z(-\pi)R_x\left(\frac{\pi}{2}\right)R_z(\pi)R_z\big( \theta_2 (\hat{\vec{\alpha}}_\text{SP})\big) R_x \left(\frac{\pi}{2}\right)R_z \big(\theta_1 (\hat{\vec{\alpha}}_\text{SP}) \big)
\label{eq::corr}
\end{align}
will result in a high-fidelity alignment of the estimated faulty (possibly mixed) initial state to the target ideal state, if the angles $\theta_1$ and $\theta_2$ are chosen such that
\begin{align*}
\theta_1\big(\hat{\vec{\alpha}}_\text{SP}\big)=-\arctan\left(\frac{\hat{\alpha}^y_\text{SP}}{\hat{\alpha}^x_\text{SP}}\right), 
\qquad 
\theta_2\big(\hat{\vec{\alpha}}_\text{SP}\big)=\arcsin\Bigg(\frac{\hat{\alpha}^z_\text{SP}}{ \big| \hat{\vec{\alpha}}_\text{SP}\big|}\Bigg)-\frac{\pi}{2}.
\end{align*}
To understand the effect of the above rotation, note that it maps a coherent error in the initial state $\rho$ to the (possibly mixed) ground state $\underline{\rho}$ in four steps:
\begin{enumerate}
    \item Rotate the state such that its projection in the $x$-$y$ plane is along the $x_+$-axis of the Bloch sphere using $R_z (\theta_1 (\hat{\vec{\alpha}}_\text{SP}))$, resulting in 
    \begin{align*}
        \rho_1&=R_z (\theta_1 (\hat{\vec{\alpha}}_\text{SP}))\rho\,R_z^\dagger (\theta_1 (\hat{\vec{\alpha}}_\text{SP}))=\frac{1}{2}\left[\mathbb{1}+\sqrt{|\hat{\vec{\alpha}}|^2-\left(\hat{\alpha}^z_\text{SP}\right)^2}\sigma_x+\hat{\alpha}^z_\text{SP}\sigma_z\right].
    \end{align*}
    \item Rotate the state onto the $x$-$y$ plane of the Bloch sphere using $R_x \!\left(\frac{\pi}{2}\right)$,
    \begin{align*}
        \rho_2&=R_x\left(\frac{\pi}{2}\right)\rho_1 R_x^\dagger\left(\frac{\pi}{2}\right) 
                =\frac{1}{2}\left[\mathbb{1}+\sqrt{|\hat{\vec{\alpha}}|^2-\left(\hat{\alpha}^z_\text{SP}\right)^2}\sigma_x-\hat{\alpha}^z_\text{SP}\sigma_y\right].
    \end{align*}
    \item Align the resulting state with the $y_-$-axis using $R_z( \theta_2 (\hat{\vec{\alpha}}_\text{SP}))$,
    \begin{align*}
        \rho_3&=R_z (\theta_2 (\hat{\vec{\alpha}}_\text{SP}))\rho_2 R_z^\dagger (\theta_2 (\hat{\vec{\alpha}}_\text{SP}))=\frac{1}{2}\left[\mathbb{1}-|\hat{\vec{\alpha}}|\sigma_y\right].
    \end{align*}
    \item Rotate the vector to align with the $z_+$-axis, finally resulting in 
    \begin{align*}
        \underline{\rho}&=\frac{1}{2}\left( \mathbb{1}+|\hat{\vec{\alpha}}|\sigma_z\right).
    \end{align*}
\end{enumerate}

In the IBM-Q devices we have considered, the values of $\hat{\alpha}^{x/y}_\text{SP}$ are small, and numerical instabilities are bound to arise in the above expression for $\theta_1$. We can avoid these by comparing the values of the parameters to the errors in the estimates, which can be expressed as $\hat{\alpha}^{x/y,(i)}_\text{SP}=\hat{\alpha}^{x/y,(i)}_{\text{SP},0}\pm\Delta\hat{\alpha}^{x/y,(i)}_\text{SP}$. Then, we can proceed as follows:
\begin{itemize}
    \item if $|\hat{\alpha}^{x,(i)}_{{\rm SP},0}| \leq 2\Delta\hat{\alpha}^{x,(i)}_{\rm SP}$, set $\hat{\alpha}^{x,(i)}_{\rm SP}=0$ and $\theta_1 = -\frac{\pi}{2}$,
    \item if $|\hat{\alpha}^{y,(i)}_{{\rm SP},0}| \leq 2\Delta\hat{\alpha}^{y,(i)}_{\rm SP}$, set $\hat{\alpha}^{y,(i)}_{\rm SP}=0$ and $\theta_1 = 0$.
\end{itemize}

\subsection{Mitigating measurement error}
\label{sub::mmit}

With the knowledge of the parameters $\hat{\alpha}^{(i)}_\text{M}$ and $\hat{\delta}_i$, it is possible to mitigate M errors from the estimates of observable expectation values. Let us consider a single-qubit measurement in $|z_\pm\rangle$ to start. In the error-free scenario, projection operators $\mathbb{P}^{(i)}_{z_\pm}$ define ideal measurements on qubit $i$, and the outcomes of such a measurement can be represented as a probability vector,
\begin{align}
    \vec{P}^{(i)}[\cdot]&\equiv \begin{bmatrix}
        \text{Tr}\big( \mathbb{P}^{(i)}_{z_+}\cdot \big)\\
        \\
        \text{Tr}\big( \mathbb{P}^{(i)}_{z_-}\cdot \big)
    \end{bmatrix}.
\end{align}
Consider now faulty measurements as in Sec.\,\ref{sub:diagonal}, Eq.\,\eqref{eq::povm1}. In order to account for the possibility of a faulty outcome, we define a new faulty probability vector, $\vec{\Pi}^{(i)}[\cdot]$, that is related to $\vec{P}^{(i)}[\cdot]$ through the invertible transformation provided by the confusion matrix $A^{(i)}$,
\begin{align}
\label{eq::Pvec}
    {\vec{\Pi}}^{(i)}[\cdot]&=\begin{bmatrix}
       \text{Tr}\left(\Pi^{(i)}_{z_+}\cdot\right)\\
        \\
        \text{Tr}\left(\Pi^{(i)}_{z_-}\cdot\right)
    \end{bmatrix}=A^{(i)}\begin{bmatrix}
        \text{Tr}\big(\mathbb{P}^{(i)}_{z_+}\cdot\big)\\
        \\
        \text{Tr}\big( \mathbb{P}^{(i)}_{z_-}\cdot\big)
    \end{bmatrix}.
\end{align}
For $\vec{\Pi}^{(i)}$ to represent the M outcomes corresponding to the parameterized diagonal POVM elements in Eq.\,\eqref{eq::povm1}, the confusion matrix $\hat{A}^{(i)}$ has the form in Eq.\,\eqref{eq::confused}, and can be constructed from the estimates of the M error parameters. The proper M outcome can be estimated through
\begin{align}
\label{eq::Mmit}
    \hat{\vec{P}}^{(i)}[\cdot]&=\left[\hat{A}^{(i)}\right]^{-1}\vec{\Pi}^{(i)}[\cdot].
\end{align}
According to our parameterization of $\hat{A}^{(i)}$, $\det{(\hat{A}^{(i)})}=\hat{\alpha}_\text{M}^{(i)}$, and $\alpha^{(i)}_\text{M}\neq 0$ in actual devices, meaning that $\hat{A}^{(i)}$ is typically invertible. Written explicitly,
\begin{align}
\label{eq::Mmitted}
    \hat{\vec{P}}^{(i)}[\cdot]&=\begin{bmatrix}
        \frac{1+\hat{\alpha}^{(i)}_\text{M}-\hat{\delta}_i}{2\hat{\alpha}^{(i)}_\text{M}} & -\frac{1-\hat{\alpha}^{(i)}_\text{M}+\hat{\delta}_i}{2\hat{\alpha}^{(i)}_\text{M}}\\
        \\
       - \frac{1-\hat{\alpha}^{(i)}_\text{M}-\hat{\delta}_i}{2\hat{\alpha}^{(i)}_\text{M}} & \frac{1+\hat{\alpha}^{(i)}_\text{M}+\hat{\delta}_i}{2\hat{\alpha}^{(i)}_\text{M}}
    \end{bmatrix}\begin{bmatrix}
      \text{Tr}\left(\Pi^{(i)}_{z_+}\cdot\right)\\
        \\
        \text{Tr}\left(\Pi^{(i)}_{z_-}\cdot\right)
    \end{bmatrix}.
\end{align}
This analysis can be extended to an $N$-qubit measurement setting. If, as we assumed, the $N$-qubit measurements are uncorrelated, the corresponding POVM elements can be expressed as tensor products, $\vec{\Pi}[\cdot]\equiv [\bigotimes^N_{i=1}\vec{\Pi}^{(i)}][\cdot]$, and the confusion matrix can be constructed as a tensor-product as well, $\hat{A}\equiv \bigotimes_{i=1}^N \hat{A}^{(i)}$. Then, the M-error-mitigated M outcomes can be obtained by inverting the confusion matrix, $\hat{\vec{P}}[\cdot]=[\hat{A}^{-1}]\vec{\Pi}[\cdot]$.

In practice, probabilities inferred from Eq.\,\eqref{eq::Mmitted} form a quasi-probability distribution, as there may exist some entries with $\hat{P}_i[\cdot]<0$ (they are, however, normalized by construction). These (quasi-)probabilities can nonetheless be used to construct unbiased estimates of observable expectation values \cite{PhysRevA.103.042605}. An efficient algorithm \cite{PRXQuantum.2.040326} to evaluate the error-mitigated M outcomes is discussed in \ref{app:readout}. {We remark that there also exist methods to reconstruct the M error-mitigated probability distribution so that it is properly normalized and positive, by using a bounded-minimization approach \cite{SmolinEfficient, BoEfficient}. This approach, however, does not provably generate an unbiased estimate for the observable expectation value.}

\subsection{Deviations from the tensor product confusion matrix}
\label{sub:crosstalk}

{As noted, our QSPAM protocol relies on the assumption that M errors are uncorrelated across qubits, allowing the full confusion matrix to be reconstructed as a tensor product of single-qubit confusion matrices. While crosstalk and coherent errors in superconducting processors can violate this assumption, the error introduced by the uncorrelated approximation is bounded and well-understood in the context of scalable mitigation techniques.}

Quantitative analyses of readout noise have shown that the uncorrelated model often provides a robust first-order approximation. For instance, \cite{PhysRevA.103.042605} compared the full confusion matrix (capturing all correlations) against the tensor product model for system sizes $N=4$ to $7$. Therein, the authors found that the total variation distance between the two models remains small, typically in the range of $0.02$ to $0.06$. Furthermore, \cite{PRXQuantum.2.040326} demonstrated that M error-mitigation methods operating on reduced subspaces remain effective even in the presence of correlated errors (e.g., those induced by increasing readout amplitudes). While it was observed that a pure tensor product model may locally {\em over-correct} in the presence of strong correlations, the overall mitigation performance was found to remain competitive. Since QSPAM employs a similar inversion logic to these established methods, we expect the bias introduced by neglecting off-diagonal terms in the multiqubit POVM to be bounded and comparable to these established benchmarks.

\bigskip

\section{Application to generating GHZ states}
\label{sec::GHZ}

To show the characterization and mitigation of SPAM errors in action, we now consider the generation of a GHZ state on a noisy multiqubit system, first in a numerical experiment, then again on an actual IBM-Q device. In both cases, we first separately characterize the SP and M errors on the system, then implement a circuit to prepare a GHZ state, and compare the observable expectation values of a target collective observable, $\braket{Z^{\otimes N}}$, without and with SPAM error-mitigation. In the latter case, we further compare the results obtained by employing standard vs. QSPAM-based protocols.
  
\subsection{Simulation of SPAM error-mitigated dynamics} 
  
For this section, we find it convenient to switch from the theoretical notation to the computational notation, namely, $z_+\mapsto 0$ and $z_-\mapsto 1$, and index the basis for an $N$-qubit Hilbert space by the bit-string corresponding to the particular state, where the set of all bit-strings corresponding to the basis states is represented by $\mathcal{B}$. For example, if $N=2$ and we have a state $\ket{10}$, we express it as $\ket{\vec{\beta}}$, where $\vec{\beta}=[1,0]$ and $\vec{\beta}\in\mathcal{B}=\{[0,0], [0,1], [1,0], [1,1]\}$ (this same notation is also used in \ref{app:readout}). The circuit used to generate a $N$-qubit GHZ state is a Hadamard operation on the first qubit, followed by chained CNOT operations on all the qubits in the system. A Lindblad simulator which takes into account realistic $T_1$ and $T_2$ noise processes, M errors as well as gate-errors is used for this simulation through the Qiskit's `fake\_provider' library \cite{qiskit2024}. In our simulated experiments, we inject SP errors through random single-qubit rotations on the ground state.

We simulate the $N$-qubit dynamics to determine the measured expectation value $\widehat{\braket{Z^{\otimes N}}}_{\rho_\text{GHZ}}$ which, ideally, should be $\braket{Z^{\otimes N}}_{\rho_\text{GHZ}}=1$ for even $N$. In reality, this expectation value will fall off as $N$ increases, due to the combined effect of the errors in the two-qubit operations, dephasing and relaxation noise processes, and SPAM errors. Notably, non-local operators such as $Z^{\otimes N}$ often represent stabilizers for quantum error-correcting codes; since the effect of the SPAM errors on the estimated expectation value is compounded with the size of the system, the improvement from SPAM mitigation becomes all the more important.

\begin{figure}[t]
    \centering
    \includegraphics[width=0.95\linewidth]{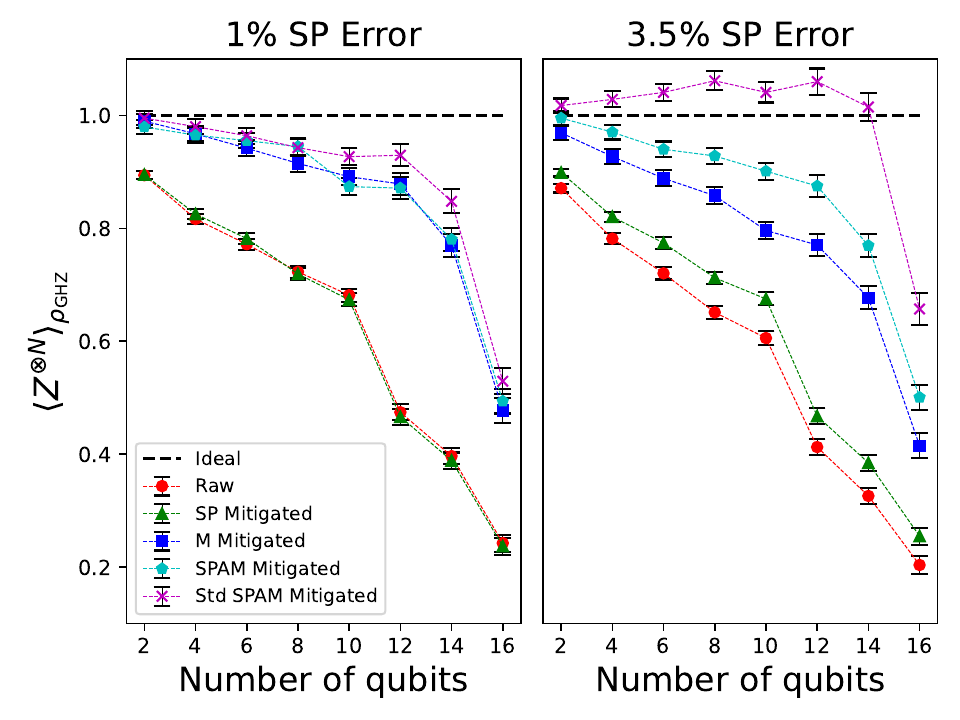}
    \caption{
    {\small
    Results for the preparation of a GHZ state with various mitigation strategies applied to the fiducial state provided by a simulated clone of the \texttt{ibm\_brisbane} device with simulated SP and M errors. The estimate of the observable expectation values is constructed over $2^{14}$ shots. The plot on the left depicts a setting where the SP error on each qubit is close to $1\%$, whereas the plot on the right has a $3.5\%$ SP error on each qubit. The different lines depict the reconstruction of the observable expectation value under different mitigation techniques. The standard characterization and mitigation approach (magenta) performs well when the SP errors are small, but significantly overestimates the value when the SP errors are appreciable. The SPAM mitigation that QSPAM generates (cyan) gives a much more accurate characterization of the expectation value.}}
    \label{fig:GHZ}
\end{figure}

The SP-error-mitigation is implemented through the corrective pulses in Eq.\,\eqref{eq::corr}, under the assumption that the fiducial $N$-qubit SP is uncorrelated. We must analyze the M error-mitigation more closely, as the confusion matrix is estimated from $N$-qubit measurements. In the presence of M errors, the expected value of the noisy estimate of $\braket{Z^{\otimes N}}_{\rho_\text{GHZ}}$ is
\begin{align}
    \widehat{\braket{Z^{\otimes N}}}_{\rho_\text{GHZ}}&=\sum_{\vec{\gamma}\in\mathcal{P}}(-1)^{\sum_i\gamma_i}\Pi_{\vec{\gamma}}[\rho],
\end{align}
where $\mathcal{P}$ is the set of outcomes for an experiment, $\Pi_{\vec{\gamma}}\equiv\bigotimes_{i=1}^N \Pi^{(i)}_{\gamma_i}$ is an element of the POVM for the $N$-qubit faulty measurements, and the single qubit POVMs are now indexed as $\Pi^{(i)}_0$ and $\Pi^{(i)}_1$. By using the inverse of the characterized confusion matrix, we can construct the M-error-mitigated expectation value by following a standard procedure inspired by \cite{PRXQuantum.2.040326} and detailed in \ref{app:readout}. This yields 
\begin{align}
    \overline{\braket{Z^{\otimes N}}}_{\rho_\text{GHZ}}&=\sum_{\vec{\beta}\in\mathcal{B}}\sum_{\vec{\gamma}\in\mathcal{P}}(-1)^{\sum_i\beta_i}\prod_{i=1}^N\left[\hat{A}^{(i)}\right]^{-1}_{\beta_i,\gamma_i}\Pi_{\vec{\gamma}}[\rho],
\end{align}
with a corresponding standard-deviation in the noisy estimate found as (see\ref{app:precision})
\begin{equation}
\sigma_{Z^{\otimes N}}\sim \frac{1}{\sqrt{\nu}} \max_{\vec{\gamma}}\sum_{\vec{\beta}\in\mathcal{B}}\prod_{i=1}^N \,\big | [A^{(i)}]^{-1}_{\beta_i,\gamma_i}\big|.\label{eq:st}
\end{equation}

In Fig.\,\ref{fig:GHZ}, we show how the estimate of the expectation value changes with different SPAM mitigation techniques. The SP mitigation operation is introduced before the GHZ state is prepared -- but after the SP noise is injected -- and the readout from the system is used to construct the expectation values. The plotted expectation values were measured without any mitigation, SP error mitigation only, M error mitigation only, QSPAM mitigation and standard SPAM mitigation. Our data clearly indicate that accounting for SP errors -- which standard characterization protocols do not -- helps avoiding over-estimating the observable expectation value as well as non-physical outcomes produced by the corresponding biased estimator. Considering the significant levels of SP errors on currently available devices (see Fig.\,\ref{fig:QSPAMbris}), not accounting for SP errors can thus lead to faulty estimates of observable expectation values. We further analyze this issue in what follows, by more precisely quantifying the deviation between representative expectation values as obtained in standard vs.\,QSPAM-based error-mitigation protocols.

\subsection{Comparison to standard SPAM error-mitigation and verification}
\label{sub::stdspam}

As discussed in Sec.\,\ref{sec:std}, standard SPAM characterization protocols estimate the product of the SPAM parameters $\alpha_\text{M}\alpha^z_\text{SP}$ and, under the assumption that the SP errors are negligible, use it to construct an approximate confusion matrix of the form given in Eq.\,\eqref{eq::confused} on each qubit; that is, if $A=\bigotimes_{i=1}^N {A}^{(i)}$ denotes the correct confusion matrix and $B=\bigotimes_{i=1}^N {B}^{(i)}$, one lets
\begin{align}
    \hat{A}^{(i)}&=\begin{bmatrix}
        \frac{1+\hat{\alpha}_\text{M}^{(i)}\hat{\alpha}^{z,(i)}_\text{SP}+\hat{\delta}_i}{2} & \frac{1-\hat{\alpha}_\text{M}^{(i)}\hat{\alpha}^{z,(i)}_\text{SP}+\hat{\delta}_i}{2}\\
        \\
        \frac{1-\hat{\alpha}_\text{M}^{(i)}\hat{\alpha}^{z,(i)}_\text{SP}-\hat{\delta}_i}{2} & \frac{1+\hat{\alpha}_\text{M}^{(i)}\hat{\alpha}^{z,(i)}_\text{SP}-\hat{\delta}_i}{2}
    \end{bmatrix} \approx 
    \begin{bmatrix}
        \frac{1+\hat{\alpha}_\text{M}^{(i)}+\hat{\delta}_i}{2} & \frac{1-\hat{\alpha}_\text{M}^{(i)}+\hat{\delta}_i}{2}\\
        \\
        \frac{1-\hat{\alpha}_\text{M}^{(i)}-\hat{\delta}_i}{2} & \frac{1+\hat{\alpha}_\text{M}^{(i)}-\hat{\delta}_i}{2}
    \end{bmatrix} = \hat{B}^{(i)}.
\end{align}
In doing so, since in reality $\hat{\alpha}_\text{M}^{(i)} \hat{\alpha}^{z,(i)}_\text{SP}< \hat{\alpha}^{(i)}_\text{M}$, the standard approach may {\em over-correct} the measured outcomes, as the results in Fig.\,\ref{fig:GHZ} reveal. This may be understood by looking at the M-error-mitigated biased estimate for the observable expectation value for a multiqubit operator $O$. If $\rho$ denotes the initially prepared (faulty) state, and $\vec{\gamma}$ ranges over the set of all reported readouts $\mathcal{P}$, we may express the estimated expectation values in the basis that diagonalizes $O$ as follows:
\begin{align}
\overline{\braket{O}}_\rho&\equiv \sum_{\vec{\gamma}\in\mathcal{P}}O(\vec{\gamma})\left[\hat{B}^{-1}\vec{{\Pi}}\right]_{\vec{\gamma}},
\end{align}
and compare it to the estimate generated using $A$, 
\begin{align}
\underline{\braket{O}}_\rho&\equiv \sum_{\vec{\gamma}\in\mathcal{P}}O(\vec{\gamma})\left[\hat{A}^{-1}\vec{\Pi}\right]_{\vec{\gamma}} .
\end{align}
The difference between the two estimates
can be upper-bounded in terms of the SP error parameters as follows:
\begin{align*}
    \left|\overline{\braket{O}}_\rho-\underline{\braket{O}}_\rho\right|& 
    =\bigg| \sum_{\vec{\gamma}\in\mathcal{P}}O(\vec{\gamma})\left[\left(\hat{B}^{-1}-\hat{A}^{-1}\right)\vec{\Pi}\right]_{\vec{\gamma}} \bigg| 
\leq\sum_{\vec{\gamma}\in\mathcal{P}}\left|\left[\left(\hat{B}^{-1}-\hat{A}^{-1}\right)\vec{\Pi}\right]_{\vec{\gamma}}\right|, 
\end{align*}
where we used the fact that the eigenvalues obey $|O(\vec{\gamma})|\leq 1$ and the triangle inequality. Since both the confusion matrices are constructed as a tensor product, we may then rewrite the above in terms of the single-qubit confusion matrices indexed over the set of qubits as follows:
\begin{align}
    \left|\overline{\braket{O}}_\rho-\underline{\braket{O}}_\rho\right|&\leq\sum_{\vec{\gamma}\in\mathcal{P}}\bigg| \bigg\{\bigotimes_{i=1}^N\left[B^{(i)}\right]^{-1}\vec{\Pi}^{(i)}-\bigotimes_{i=1}^N\left[A^{(i)}\right]^{-1}\vec{\Pi}^{(i)}\bigg\}_{\vec{\gamma}}\bigg|.
    \label{eq:ub}
\end{align}
Note that $\vec{\Pi}^{(i)}$ is the marginal distribution for a single qubit in an uncorrelated multi-qubit readout, and represents a binomial distribution with parameter $p=\vec{\Pi}^{(i)}_0$. Writing out the matrix multiplication explicitly, and using $\vec{\Pi}^{(i)}_1=1-\vec{\Pi}^{(i)}_0$, we obtain 
\begin{align}
\label{eq::tensor1}
    \left[B^{(i)}\right]^{-1}\!\vec{\Pi}^{(i)}=\!\begin{bmatrix}
        \frac{1}{2}-\left(\frac{1-2\Pi_o^{(i)}+\delta_i}{2\alpha_\text{M}^{(i)}}\right)\\
        \frac{1}{2}+\left(\frac{1-2\Pi_o^{(i)}+\delta_i}{2\alpha_\text{M}^{(i)}}\right)
    \end{bmatrix}, \quad 
    \left[A^{(i)}\right]^{-1}\!\vec{\Pi}^{(i)}=\!\begin{bmatrix}
        \frac{1}{2}-\left(\frac{1-2\Pi_o^{(i)}+\delta_i}{2\alpha_\text{M}^{(i)}\alpha_\text{SP}^{z,(i)}}\right)\\
        \frac{1}{2}+\left(\frac{1-2\Pi_o^{(i)}+\delta_i}{2\alpha_\text{M}^{(i)}\alpha_\text{SP}^{z,(i)}}\right)
    \end{bmatrix}.
\end{align}
Recall that $\vec{\gamma}$ is a binary vector of length $2^N$, whose $j$th component indexes the marginal distribution vector ($\vec{\Pi}^{(i)}_0$ or $\vec{\Pi}^{(i)}_1$). This allows us to simplify Eq.\,\eqref{eq:ub} as 
\begin{align}
    \left|\overline{\braket{O}}_\rho-\underline{\braket{O}}_\rho\right| &\leq\sum_{\vec{\gamma}\in\mathcal{P}} \, \left| \prod_{i=1}^N\left\{\left[B^{(i)}\right]^{-1}\vec{\Pi}^{(i)}\right\}_{\gamma_i}-\prod_{i=1}^N\left\{\left[A^{(i)}\right]^{-1}\vec{\Pi}^{(i)}\right\}_{\gamma_i}\right|.
\end{align}
Substituting the expressions from Eq.\,\eqref{eq::tensor1}, this yields 
\begin{align}
\label{eq::prodDiff}
    \left|\overline{\braket{O}}_\rho-\underline{\braket{O}}_\rho\right|& \leq\sum_{\vec{\gamma}\in\mathcal{P}} \bigg|\prod_{i=1}^N \bigg\{\frac{1}{2}+(-1)^{\gamma_i+1} \bigg(\frac{1-2\Pi_o^{(i)}+\delta_i}{2\alpha_\text{M}^{(i)}}\bigg) \bigg\}\notag\\
    &\qquad- \prod_{i=1}^N \bigg\{\frac{1}{2}+(-1)^{\gamma_i+1}\bigg(\frac{1-2\Pi_o^{(i)}+\delta_i}{2\alpha_\text{M}^{(i)}\alpha_\text{SP}^{z,(i)}}\bigg)\bigg\}\bigg|.
\end{align}

We can now use the inequality, $|\prod_{i=1}^N a_i-\prod_{i=1}^N b_i|\leq\sum_{i=1}^N|a_i-b_i|$ if $|a_i|,|b_i|\leq 1$, to further simplify the above expression. If the SPAM errors are {\em weak}, such that, for all $i$, 
\begin{align}
    \bigg| \frac{1}{2}\pm\bigg(\frac{1-2\Pi_o^{(i)}+\delta_i}{2\alpha_\text{M}^{(i)}}\bigg)\bigg|\leq 1, \qquad 
    \bigg|\frac{1}{2}\pm\bigg(\frac{1-2\Pi_o^{(i)}+\delta_i}{2\alpha_\text{M}^{(i)}\alpha_\text{SP}^{z,(i)}}\bigg)\bigg|\leq 1,
\end{align}
we may finally write 
\begin{align}
    \left|\overline{\braket{O}}_\rho-\underline{\braket{O}}_\rho\right|& \leq\sum_{\vec{\gamma}\in\mathcal{P}}\sum_{i=1}^N \bigg|\frac{1}{\alpha^{z,(i)}_\text{SP}}-1\bigg | \bigg |\frac{1-2\Pi^{(i)}_0+\delta_i}{2\alpha_\text{M}^{(i)}}\bigg | \notag \\
\label{eq::diffScale}
   & 
   \leq|\mathcal{P}|\sum_{i=1}^N\bigg|\frac{1}{\alpha^{z,(i)}_\text{SP}}-1\bigg |\bigg | \frac{1-2\Pi^{(i)}_0+\delta_i}{2\alpha_\text{M}^{(i)}}\bigg|.
\end{align}
{This bound implies that, in the regime where SPAM errors are small, the bias in the observable estimator scales \emph{linearly} with both the system size $N$ and the magnitude of the SP errors, $1-\alpha_\text{SP}^{z,(i)}$ (to the first order). In fact, a bound similar to the one in Eq.\,\eqref{eq::diffScale} may also be derived by relaxing the assumption that SPAM errors are small (see \ref{app:largesp}), in which case the difference between the two estimates diverges as a sum of the inverse powers of the SP parameters. Either way, an unbiased estimate for the observable expectation value may only be constructed if the SP errors are taken into account for all qubits measured, emphasizing the importance of separate SP characterization for scalable quantum processors.}

\begin{figure}
    \centering
    \includegraphics[width=.8\linewidth]{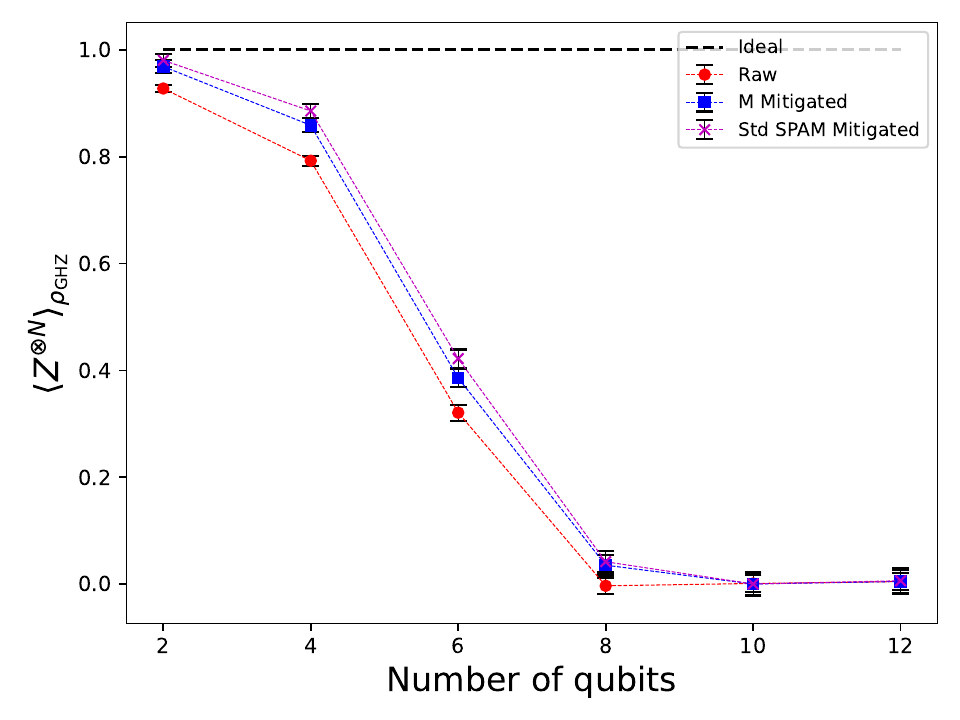}
    \vspace*{-3mm}
    \caption{
    {\small Preparation of a GHZ state and subsequent estimation of the observable $\braket{Z^{\otimes N}}_{\rho_\text{GHZ}}$, with various mitigation strategies applied to the fiducial state provided by the \texttt{ibm\_sherbrooke} device. The estimate of the observable expectation values is constructed over $2^{14}$ shots on a linearly connected set of up to $12$ qubits. The magenta line shows the result of standard M error mitigation, where SP is assumed to be perfect, the blue line shows the result of M error mitigation using the results of the QSPAM protocol, and the red line shows the raw result where no M error mitigation strategy is applied. The plot illustrates how assuming that the SP is perfect leads to an overestimate of the observable.}}
    \label{fig:GHZReal}
\end{figure}

Fig.\,\ref{fig:GHZReal} demonstrates the behavior of this difference on an actual IBM-Q device, where a GHZ state is prepared by selecting a linearly connected set of qubits of size $N\in\{2,4,6,8,10,12\}$. The raw measurement of this expectation value is under-estimated, as expected, due to unaddressed SPAM errors. When SPAM mitigation is carried out using the above $A$ and $B$ matrices, we see a statistically significant distance between the corresponding expectation values $\underline{\braket{Z^{\otimes N}}}_{\rho_\text{GHZ}}$ and $\overline{\braket{Z^{\otimes N}}}_{\rho_\text{GHZ}}$. Since the expectation values are estimated over the same dataset, the only source of this disagreement can be the unaddressed SP errors. Unlike the simulated setting shown in Fig.\,\ref{fig:GHZ}, the expectation values are seen to fall to zero -- due to gate and environmental noise -- more rapidly for $N\geq 10$, and the mitigation strategies no longer produce results any different from the raw data.

\section{Discussion and outlook}
\label{sec::discussion}

In this paper, we have presented the QSPAM protocol, which enables users to characterize uncorrelated SP and M errors independently on NISQ-era and early fault-tolerant devices using only single-qubit operations and measurements. QSPAM does not require ancillae or two-qubit operations, which are generally time-consuming and can be lower in fidelity than single-qubit gates by an order of magnitude. As a consequence, for platforms where non-destructive measurements are available, QSPAM is resource-efficient and highly parallelizable compared to existing protocols. The depth of the circuit required for a parallel QSPAM characterization is also constant in system size and, therefore, does not suffer from compounding noise effects that emerge from deeper circuits \cite{PhysRevA.106.012439,Laflamme2021}. The precision of the characterization scales in the number of shots as $\nu^{-1/2}$, with a measurement overhead that is upper-bounded by the matrix elements of the confusion matrix $A$, which are entirely device-dependent. These advantages make the QSPAM protocol a very efficient tool; in particular, it can be readily integrated into the existing suite of SPAM characterization and mitigation strategies deployed on a variety of NISQ-era devices.

An important question concerns the extent to which our assumptions of diagonal POVM elements and efficient measurements are justified on real hardware. Ref. \cite{pereira_parallel_2023} performs model-free measurement tomography on IBM's older \texttt{ibm\_perth} device, which provides independent validation through reconstruction of the full Choi matrices of the measurement process. In this study, the off-diagonal POVM elements are found to be negligible (below $10^{-2}$) compared to the diagonal assignment errors across all characterized qubits. The Choi matrices corresponding to the measurement outcomes are rank-1 to good approximation, meaning that the measurement can be considered to be efficient, consistent with our assumptions. An experimental self-consistency check based on three consecutive measurements, in which the model predictions are validated against the observed statistics of a third readout, would provide a direct, device-level test of the underlying measurement model assumptions. We view this as a worthwhile validation strategy for future work on next-generation devices.

The QSPAM protocol that we have presented also assumes that readouts across the device are uncorrelated and, consequently, relies on a tensor product POVM. It has been shown that a tensor product POVM can provide an accurate approximation for readouts in a variety of current NISQ devices \cite{PRXQuantum.2.040326,Zoltan,Geller_2021,arrasmith2023developmentdemonstrationefficientreadout}. However, if significant correlations are present in readout across the device (such as the ones that may arise out of a multiplexed measurement) the tensor-product POVM-assumption over-corrects the estimated expectation values \cite{PRXQuantum.2.040326, PRXQuantum.6.010307}. Generalizing the QSPAM protocols to settings with correlated readout errors remains an interesting area for future theoretical and experimental work.

\smallskip

\section*{Acknowledgments}

We thank Mattias Fitzpatrick, Juan Sebastian Salcedo-Gallo, and Chris Wang for valuable input and discussions. It is also a pleasure to thank Gerardo Paz-Silva for calling attention to the issue of gauge freedom and Victor Colussi for help in elucidating important aspects of the measurement model we consider.  We acknowledge the use of IBM Quantum services for this work. The views expressed are those of the authors, and do not reflect the official policy or position of IBM or the IBM Quantum team. We also acknowledge the use of resources of the Oak Ridge Leadership Computing Facility, which is a Department of Energy (DOE) Office of Science User Facility supported under Contract DE-AC05-00OR22725. 
L.M.N. was supported in part by the DOE, Office of Science, Office of Advanced Scientific Computing, Research, Accelerated Research in Quantum Computing under Award Number DE-SC0020316 and DE-SC0025509. 
Work at Dartmouth was also supported by the DOE,  Office of Science, Office of Advanced Scientific Computing Research, Accelerated Research in Quantum Computing, under Award Number DE-SC0020316, and by the U.S. Army Research Office, through the multidisciplinary Research Program of the University Research Initiative (MURI) Grant No.\,W911NF-18-1-0218, and through Grant No.\,W911NF-22-1-0004.

\smallskip

\appendix

\section{sQSPAM protocol with faulty single-qubit gates}
\label{app:FG} 

It is straightforward to modify the sQSPAM protocol in the presence of faulty single-qubit gates. Suppose, in particular, that the $X$-gate and the $R_x(\frac{\pi}{2})$ do not execute proper rotations; then, we can introduce the parameters corresponding to the faulty rotations $\alpha'^z_\text{SP}$, $\alpha'^x_\text{SP}$, $\alpha'^y_\text{SP}$, and an additional experiment as described below:
\begin{align}
    \mathbb{P}_{z_+\rightarrow z_-}&=\frac{1}{2}\left(1-\alpha_\text{M}\alpha_\text{SP}^z-\delta\right),\\
    \mathbb{P}_{z_-\rightarrow z_+}&=\frac{1}{2}\left(1-\alpha_\text{M}\alpha_\text{SP}'^z+\delta\right),\\
    \mathbb{P}_{x_+\rightarrow z_-}&=\frac{1}{2}\left(1-\alpha_\text{M}\alpha_\text{SP}'^x-\delta\right),\\
    \mathbb{P}_{y_+\rightarrow z_-}&=\frac{1}{2}\left(1-\alpha_\text{M}\alpha_\text{SP}'^y-\delta\right),\\
    \mathbb{P}_{z_+\rightarrow z_+\rightarrow z_+}&=\frac{\alpha_\text{M}^2+2\alpha_\text{M}\alpha_\text{SP}^z(1+\delta)+(1+\delta)^2}{2(1+\alpha_\text{M}\alpha_\text{SP}^z+\delta)},\\
    \mathbb{P}_{z_-\rightarrow z_+\rightarrow z_+}&=\frac{\alpha_\text{M}^2-2\alpha_\text{M}\alpha_\text{SP}'^z(1+\delta)+(1+\delta)^2}{2(1-\alpha_\text{M}\alpha_\text{SP}'^z+\delta)}.
\end{align}
These equations may be solved for the SP parameter $\alpha_\text{SP}^z$ and the M parameters $\alpha_\text{M}$, $\delta$. 

As we noted in the main text, gauge ambiguity emerges in such a setting, similar to GST \cite{Nielsen2021gatesettomography}. In such a situation, the preferred frame is no longer obvious, and there is freedom in what we choose to gauge-optimize. If there is confidence that the gate-fidelities are better than M and SP fidelities, we may search for a gauge that maximizes the gate fidelities and characterize the SP and M parameters in this particular gauge. Once a particular gauge is found, the estimated parameters $\alpha'^x_\text{SP}$, $\alpha'^y_\text{SP}$ may be mapped back to $\alpha^x_\text{SP}$, $\alpha^y_\text{SP}$ to complete the characterization.

\section{Analytical solution for SPAM parameters}
\label{app:analyticSPAM}

{
We derive analytical expressions for the SPAM parameters by inverting the system of equations governing the experimental probabilities defined in Eqs.\,\eqref{eq::p0to1F}-\eqref{eq::p1to0to0piF}. Let us define the auxiliary sum variables $\Sigma_\pm$ corresponding to the repeated measurement experiments:
\begin{align*}
    \Sigma_\pm
    \equiv \mathbb{P}^{\theta_z=0}_{z_\pm\rightarrow z_+\rightarrow z_+} \!+ \mathbb{P}^{\theta_z=\pi}_{z_\pm\rightarrow z_+\rightarrow z_+} 
    = \frac{\alpha_\text{M}^2(1-\epsilon) \pm 2\alpha_\text{M}\alpha_\text{SP}^z(1+\delta)+(1+\delta)^2(1+\epsilon)}{(1 \pm \alpha_\text{M}\alpha_\text{SP}^z+\delta)(1+\epsilon)}.
\end{align*}
Let $\kappa \equiv \alpha_\text{M}\alpha_\text{SP}^z$. From the single-measurement probabilities $\mathbb{P}_{z_+\rightarrow z_-}$ and $\mathbb{P}_{z_-\rightarrow z_+}$, we can immediately isolate the asymmetry parameter $\delta$ and the product $\kappa$:
\begin{align*}
    \delta = \mathbb{P}_{z_-\rightarrow z_+} - \mathbb{P}_{z_+\rightarrow z_-}, \qquad 
    \kappa = 1 - \left(\mathbb{P}_{z_+\rightarrow z_-} + \mathbb{P}_{z_-\rightarrow z_+}\right).
\end{align*}
With $\delta$ and $\kappa$ determined, we can next define the auxiliary quantities $Q_\pm$ to simplify the non-linear components of the system:
\begin{align*}
    Q_+ &\equiv \Sigma_+(1-\mathbb{P}_{z_+\rightarrow z_-}) + \Sigma_-\mathbb{P}_{z_-\rightarrow z_+}, \\
    Q_- &\equiv \Sigma_+(1-\mathbb{P}_{z_+\rightarrow z_-}) - \Sigma_-\mathbb{P}_{z_-\rightarrow z_+}.
\end{align*}
Substituting $\delta$ and $\kappa$ into the expressions for $\Sigma_\pm$, we can solve for the non-diagonality parameter,
 $$   \epsilon = \frac{2\kappa(1+\delta)}{Q_-}.$$
Finally, using the derived value of $\epsilon$, we extract the measurement fidelity $\alpha_\text{M}$ and subsequently the SP fidelity $\alpha_\text{SP}^z$:
\begin{align*}
    \alpha_\text{M} = \sqrt{\frac{\big[Q_+-\left(1+\delta\right)^2\big]\kappa\left(1+\delta\right)}{Q_- - \kappa(1+\delta)}}, \qquad 
    \alpha^z_\text{SP} = \frac{\kappa}{\alpha_\text{M}}.
\end{align*}
This analytical solution provides a direct method to estimate the parameters, which serves as a robust initial guess for the weighted numerical minimization described in the main text. }

\bigskip

\section{Constructing measurement-error mitigated readouts}
\label{app:readout}

In the main text we presented an inversion problem to obtain the M-error free estimates of observable expectation values. Utilizing the tensor-product structure of the confusion matrix $A$, it is easy to construct the inverse of the estimate as $\hat{A}^{-1}=\bigotimes_{i=1}^N [\hat{A}^{(i)}]^{-1}$. However, this matrix has $2^{2N}$ elements, as there are $2^N$ classical registers for $N$ qubits. This means that the matrix multiplication involves $2^{2N}$ operations, which becomes infeasible as $N$ grows.

Real systems bypass this limitation by constructing a readout register consisting of {\em non-zero readouts} only, and this is typically much smaller than $2^{N}$. Let us call this readout register $\vec{\Pi}$ and index it by bit-strings corresponding to the POVM outcomes where $z_+\mapsto 0$ and $z_-\mapsto 1$. As an example, for $N=2$, $\vec{\Pi}$ is indexed by $\{[0,0], [0,1], [1,0], [1,1]\}$, such that
\begin{align*}    
\Pi_{0,0}=\left[\Pi^{(1)}_0\otimes\Pi^{(2)}_0\right][\rho],\\   \Pi_{0,1}=\left[\Pi^{(1)}_0\otimes\Pi^{(2)}_1\right][\rho],\\    \Pi_{1,0}=\left[\Pi^{(1)}_1\otimes\Pi^{(2)}_0\right][\rho],\\    \Pi_{1,1}=\left[\Pi^{(1)}_1\otimes\Pi^{(2)}_1\right][\rho].
\end{align*}
An experiment provides results in $\vec{\Pi}$ as the output, that only contains the non-zero readouts; therefore, if we can set up an inversion problem that is indexed by the present bit-strings, we can reduce the complexity of this task. Following a procedure similar to the one laid out in \cite{PRXQuantum.2.040326}, for $N$-qubits there are $2^N$ bit-strings of length $N$ that form a set that we will denote by $\mathcal{B}$. The corrected readout element $\mathbb{P}_{\vec{\beta}}$ for some $\vec{\beta}\in\mathcal{B}$ can be expressed as
\begin{align}
\hat{\mathbb{P}}_{\vec{\beta}}&=\sum_{\vec{\gamma}\in\mathcal{P}}\prod_{i=1}^N\left[\hat{A}^{(i)}\right]^{-1}_{\beta_i,\gamma_i}\Pi_{\vec{\gamma}},
\end{align}
where $\mathcal{P}$ is the set of bit-strings returned by the device, whose typical size $|\mathcal{P}|\ll 2^N$ as $N\gg 1$. The product in this expression is between $N$ matrices of size $2\times 2$, and therefore the complexity of this evaluation can be tightly bounded as $\hat{\mathbb{P}}_{\vec{\beta}}\in\Theta(N|\mathcal{P}|)$ and evaluating the set $\{\hat{\mathbb{P}}_{\vec{\beta}};\forall\vec{\beta}\in\mathcal{B}\}\in\Theta(N2^{N}|\mathcal{P}|)$. The positivity condition on the probabilities requires that
\begin{align*}
\prod_{i=1}^N\left[\hat{A}^{(i)}\right]^{-1}_{\beta_i,\eta_i}\geq 0, \qquad \forall \vec{\beta},\vec{\eta}\in\mathcal{B}.
\end{align*}
This condition need not be satisfied by experimentally estimated confusion matrices, because of small deviations between the ``true'' value and the estimated value of the M error parameters. As a consequence, the set $\{\hat{\mathbb{P}}_{\vec{\beta}}\}_{\vec{\beta}\in\mathcal{B}}$ does in general yield a quasiprobability distribution that is normalized but {\em not} strictly positive \cite{PhysRevA.100.052315,PRXQuantum.6.010307}. The estimate of an expectation value constructed from this quasiprobability distribution can still be used to obtain an unbiased estimator for the target expectation value \cite{PhysRevA.103.042605}. Specifically, for an observable $O$, we can express the biased estimate of the expectation value by measuring in the eigenbasis for $O$: 
\begin{align*}
\widehat{\braket{O}}&=\sum_{\vec{\gamma}\in\mathcal{P}}O(\vec{\gamma})\Pi_{\vec{\gamma}},
\end{align*}
where $O(\vec{\gamma})$ denoted the eigenvalue corresponding to eigenvector $\ket{\vec{\gamma}}$. The corresponding SPAM-mitigated unbiased estimate for the observable expectation value is then 
\begin{align*}
\overline{\braket{O}}&=\sum_{\vec{\beta}\in\mathcal{B}}\sum_{\vec{\gamma}\in\mathcal{P}}O(\vec{\beta})\prod_{i=1}^{\dim\vec{\gamma}}\left[\hat{A}^{(i)}\right]^{-1}_{\beta_i,\gamma_i}\Pi_{\vec{\gamma}}.
\end{align*}

\section{Errors in measurement-error mitigated readouts}
\label{app:precision}

The error in the SPAM-mitigated estimate of the observable expectation value, quantified in term of the standard deviation 
$\sigma_{O}$, can be calculated by approximating the multinomial distribution corresponding to $\hat{\vec{\Pi}}$ as a multivariate normal distribution, with a standard deviation  $\sqrt{\frac{\Pi_{\vec{\gamma}}(1-\Pi_{\vec{\gamma}})}{\nu}}\approx \sigma_{\vec{\gamma}}$ when $\nu\gg1$. This error can then be propagated into the estimate for $\overline{\braket{O}}$, yielding the variance
\begin{align*}
\sigma_{O}^2&=\sum_{\vec{\gamma}\in\mathcal{P}}\left[O(\vec{\gamma})\right]^2\sigma^2_{\vec{\gamma}} \bigg[\sum_{\vec{\beta}\in\mathcal{B}}\prod_{i=1}^{\dim \vec{\gamma}}\left[A^{(i)}\right]^{-1}_{\beta_i,\gamma_i} \bigg]^2.
\end{align*}
In the ideal case, the confusion matrix is just the identity and $\sigma_{O}\sim \nu^{-\frac{1}{2}}$, because $\sigma_{\vec{\gamma}}\sim \nu^{-\frac{1}{2}}$ and $|O(\vec{\gamma})|\leq 1,\forall\vec\gamma\in\mathcal{P}$. The M-error mitigation protocol introduces an overhead and, as a result, the upper bound on the variance of the estimate increases to $\sigma_{O}\leq \nu^{-\frac{1}{2}}\max_{\vec{\gamma}}\sum_{\vec{\beta}\in\mathcal{B}}\prod_{i=1}^N|[A^{(i)}]^{-1}_{\beta_i,\gamma_i}|$, in line with \cite{PhysRevA.103.042605} and as claimed in Eq.\,\eqref{eq:st}.

\section{Comparative M-error mitigation with large SP errors}
\label{app:largesp}

In Sec.\,\ref{sec::GHZ} we assumed the SPAM errors to be small to obtain a simple scaling form in terms of the SP parameters. If the SPAM errors are not small, however, we may still obtain an upper-bound by using a more general form of the inequality for the absolute difference of products relationships, namely,
\begin{align*}
\left| \prod_{i=1}^{N} a_i - \prod_{i=1}^{N} b_i \right| \le \sum_{k=1}^{N} \big| a_k - b_k \big| \bigg( \prod_{j=1}^{k-1} |b_j| \bigg) \bigg( \prod_{j=k+1}^{N} |a_j| \bigg).
\end{align*}
By using the above in Eq.\,\eqref{eq::prodDiff}, we obtain
\begin{align*}
\bigg|\overline{\braket{O}}_\rho-\underline{\braket{O}}_\rho \bigg| &\leq \sum_{k=1}^{N} \bigg| \frac{1}{\alpha_\text{SP}^{z,(k)}}-1\bigg| \bigg| \frac{1 - 2\Pi_0^{(k)} + \delta_k}{2\alpha_\text{M}^{(k)}} \bigg| \nonumber \\
& \qquad \qquad \times \prod_{j=1}^{k-1} \bigg| \frac{1}{2} + (-1)^{\beta_j+1} \bigg( \frac{1 - 2\Pi_0^{(j)} + \delta_j}{2\alpha_\text{M}^{(j)} \alpha_\text{SP}^{z,(j)}} \bigg) \bigg| \nonumber \\
& \qquad \qquad \times \prod_{j'=k+1}^{N} \bigg| \frac{1}{2} + (-1)^{\beta_{j'}+1} \bigg( \frac{1 - 2\Pi_0^{(j')} + \delta_{j'}}{2\alpha_\text{M}^{(j')}} \bigg) \bigg|.
\end{align*}
This tighter bound still scales similar to Eq.\,\eqref{eq::diffScale} in $\alpha_\text{SP}^{z,(j)}$.

\smallskip

\section*{References}

\bibliographystyle{bibstyle_v4}
\bibliography{main_final}

\providecommand{\newblock}{}
\begin{thebibliography}{10}
\expandafter\ifx\csname url\endcsname\relax
  \def\url#1{{\tt #1}}\fi
\expandafter\ifx\csname urlprefix\endcsname\relax\def\urlprefix{URL }\fi
\providecommand{\href}[2]{#1}  
\providecommand{\eprint}[2][arXiv]{#1:\linebreak[0]#2}

\bibitem{Degen2017}
Degen C~L, Reinhard F and Cappellaro P 2017 Quantum sensing {\em Rev. Mod.
  Phys.\/} {\bf 89} 035002

\bibitem{Felix2018}
Beaudoin F, Norris L~M and Viola L 2018 Ramsey interferometry in correlated
  quantum noise environments {\em Phys. Rev. A\/} {\bf 98} 020102

\bibitem{riberi2023}
Riberi F, Paz-Silva G~A and Viola L 2023 Nearly {H}eisenberg-limited
  noise-unbiased frequency estimation by tailored sensor design {\em Phys. Rev.
  A\/} {\bf 108} 042419

\bibitem{riberi2025}
Riberi F and Viola L 2025 Optimal asymptotic precision bounds for nonlinear
  quantum metrology under collective dephasing {\em APL Quantum\/} {\bf 2}
  026111

\bibitem{Datta}
Datta A, Zhang L, Thomas-Peter N {\em et~al\/} 2011 Quantum metrology with
  imperfect states and detectors {\em Phys. Rev. A\/} {\bf 83} 063836

\bibitem{Len2022}
Len Y~L, Gefen T, Retzker A {\em et~al\/} 2022 Quantum metrology with imperfect
  measurements {\em Nature Commun.\/} {\bf 13} 6971

\bibitem{Szankowski_2017}
Szańkowski P, Ramon G, Krzywda J {\em et~al\/} 2017 Environmental noise
  spectroscopy with qubits subjected to dynamical decoupling {\em J. Phys.:
  Cond. Matter\/} {\bf 29} 333001

\bibitem{PhysRevA.95.022121}
Paz-Silva G~A, Norris L~M and Viola L 2017 Multiqubit spectroscopy of
  {G}aussian quantum noise {\em Phys. Rev. A\/} {\bf 95} 022121

\bibitem{PazSilvaFrames}
Chalermpusitarak T, Tonekaboni B, Wang Y {\em et~al\/} 2021 Frame-based
  filter-function formalism for quantum characterization and control {\em PRX
  Quantum\/} {\bf 2} 030315

\bibitem{PhysRevApplied.22.024074}
Khan M~Q, Dong W, Norris L~M {\em et~al\/} 2024 Multiaxis quantum noise
  spectroscopy robust to errors in state preparation and measurement {\em Phys.
  Rev. Appl.\/} {\bf 22} 024074

\bibitem{Skinner2019}
Skinner B, Ruhman J and Nahum A 2019 Measurement-induced phase transitions in
  the dynamics of entanglement {\em Phys. Rev. X\/} {\bf 9} 031009

\bibitem{Garratt2023}
Garratt S~J, Weinstein Z and Altman E 2023 Measurements conspire nonlocally to
  restructure critical quantum states {\em Phys. Rev. X\/} {\bf 13} 021026

\bibitem{Murciano2023}
Murciano S, Sala P, Liu Y {\em et~al\/} 2023 Measurement-altered {I}sing
  quantum criticality {\em Phys. Rev. X\/} {\bf 13} 041042

\bibitem{PhysRevX.10.041020}
Gullans M~J and Huse D~A 2020 Dynamical purification phase transition induced
  by quantum measurements {\em Phys. Rev. X\/} {\bf 10} 041020

\bibitem{Blunt2023}
Blunt N~S, Caune L, Izsák R {\em et~al\/} 2023 Statistical phase estimation
  and error mitigation on a superconducting quantum processor {\em PRX
  Quantum\/} {\bf 4} 040341

\bibitem{IBM2025}
Kang H, Kam J~F, Mooney G~J {\em et~al\/} 2025 Teleporting two-qubit
  entanglement across 19 qubits on a superconducting quantum computer {\em
  Phys. Rev. Appl.\/} {\bf 23} 014057

\bibitem{Livingston2022}
Livingston W~P, Blok M~S, Flurin E {\em et~al\/} 2022 Experimental
  demonstration of continuous quantum error correction {\em Nat. Commun.\/}
  {\bf 13} 2307

\bibitem{Acharya2025}
Acharya R, Abanin D~A, Aghababaie-Beni L {\em et~al\/} 2025 Quantum error
  correction below the surface code threshold {\em Nature\/} {\bf 638} 920

\bibitem{Quantinuum2022}
An F~A, Ransford A, Schaffer A {\em et~al\/} 2022 High fidelity state
  preparation and measurement of ion hyperfine qubits with ${I}>\frac{1}{2}$
  {\em Phys. Rev. Lett.\/} {\bf 129} 130501

\bibitem{Laflamme2021}
Lin J, Wallman J~J, Hincks I {\em et~al\/} 2021 Independent state and
  measurement characterization for quantum computers {\em Phys. Rev. Res.\/}
  {\bf 3} 033285

\bibitem{Klimov}
Bengtsson A, Opremcak A, Khezri M, Sank D, Bourassa A, Satzinger K~J, Hong S,
  Erickson C, Lester B~J, Miao K~C, Korotkov A~N, Kelly J, Chen Z and Klimov
  P~V 2024 Model-based optimization of superconducting qubit readout {\em Phys.
  Rev. Lett.\/} {\bf 132} 100603

\bibitem{Blumoff2022}
Blumoff J~Z, Pan A~S, Keating T~E {\em et~al\/} 2022 Fast and high-fidelity
  state preparation and measurement in triple-quantum-dot spin qubits {\em PRX
  Quantum\/} {\bf 3} 010352

\bibitem{Takeda2024}
Takeda K, Noiri A, Nakajima T {\em et~al\/} 2024 Rapid single-shot parity spin
  readout in a silicon double quantum dot with fidelity exceeding 99{\%} {\em
  npj Quantum Information\/} {\bf 10} 22

\bibitem{Endres2023}
Scholl P, Shaw A~L, Tsai R~B~S, Finkelstein R, Choi J and Endres M 2023 Erasure
  conversion in a high-fidelity {R}ydberg quantum simulator {\em Nature\/} {\bf
  622} 273--

\bibitem{Bryce2024}
Chen T, Huang C, Velkovsky I, Hazzard K~R~A, Covey J~P and Gadway B 2024
  Strongly interacting {R}ydberg atoms in synthetic dimensions with a magnetic
  flux {\em Nat. Commun.\/} {\bf 15} 2675

\bibitem{Sandia}
Hashim A 2024 A practical introduction to benchmarking and characterization of
  quantum computers {\em arXiv:2408.12064 [quant-ph]\/}

\bibitem{CompressedGST}
Brieger R, Roth I and Kliesch M 2023 Compressive gate set tomography {\em PRX
  Quantum\/} {\bf 4} 010325

\bibitem{NonMarkovianGST}
Li Z, Zheng C, Meng F {\em et~al\/} 2024 Non-{M}arkovian quantum gate set
  tomography {\em Quantum Sci. Tech.\/} {\bf 9} 035027

\bibitem{White2020}
White G~A~L, Hill C~D, Pollock F~A {\em et~al\/} 2020 Demonstration of
  non-{M}arkovian process characterisation and control on a quantum processor
  {\em Nat. Commun.\/} {\bf 11} 6301

\bibitem{hurant2024}
Hurant T, Sun K, Jia Z {\em et~al\/} 2024 Few-shot, robust calibration of
  single qubit gates using {B}ayesian robust phase estimation {\em
  arXiv:2407.18339 [quant-ph]\/}

\bibitem{FastBT}
Su R~Y, Huang J~Y, Stuyck N~D {\em et~al\/} 2025 Characterizing non-{M}arkovian
  quantum processes by fast {B}ayesian tomography {\em Phys. Rev. A\/} {\bf
  111} 052425

\bibitem{QEM}
Cai Z, Babbush R, Benjamin S~C, Endo S, Huggins W~J, Li Y, McClean J~R and
  O'Brien T~E 2023 Quantum error mitigation {\em Rev. Mod. Phys.\/} {\bf 95}
  045005

\bibitem{Gupta2024}
Gupta R~S, van~den Berg E, Takita M {\em et~al\/} 2024 Probabilistic error
  cancellation for dynamic quantum circuits {\em Phys. Rev. A\/} {\bf 109}
  062617

\bibitem{vandenBerg2023}
van~den Berg E, Minev Z~K, Kandala A {\em et~al\/} 2023 Probabilistic error
  cancellation with sparse {P}auli-{L}indblad models on noisy quantum
  processors {\em Nat. Phys.\/} {\bf 19} 1116

\bibitem{Uzdin}
Santos J~P and Uzdin R 2025 Drift-resilient mid-circuit measurement and state
  preparation error mitigation for dynamic circuits {\em arXiv:2506.11270
  [quant-ph]\/}

\bibitem{PhysRevA.103.042605}
Bravyi S, Sheldon S, Kandala A {\em et~al\/} 2021 Mitigating measurement errors
  in multiqubit experiments {\em Phys. Rev. A\/} {\bf 103} 042605

\bibitem{PRXQuantum.2.040326}
Nation P~D, Kang H, Sundaresan N {\em et~al\/} 2021 Scalable mitigation of
  measurement errors on quantum computers {\em PRX Quantum\/} {\bf 2} 040326

\bibitem{Ivashkov2024}
Ivashkov P, Uchehara G, Jiang L {\em et~al\/} 2024 High-fidelity, multiqubit
  generalized measurements with dynamic circuits {\em PRX Quantum\/} {\bf 5}
  030315

\bibitem{PhysRevA.100.052315}
Chen Y, Farahzad M, Yoo S {\em et~al\/} 2019 Detector tomography on {IBM}
  quantum computers and mitigation of an imperfect measurement {\em Phys. Rev.
  A\/} {\bf 100} 052315

\bibitem{DiGiovanni}
Di~Giovanni A, Aasen A~S, Lisenfeld J, G\"arttner M, Rotzinger H and Ustinov
  A~V 2025 Benchmarking the quality of multiplexed qubit readout beyond
  assignment fidelity {\em Phys. Rev. Appl.\/} {\bf 24} 044043

\bibitem{Yu2025}
Yu H and Wei T 2025 Efficient separate quantification of state preparation
  errors and measurement errors on quantum computers and their mitigation {\em
  Quantum\/} {\bf 9} 1724

\bibitem{Mattias}
Landa H, Meirom D, Kanazawa N {\em et~al\/} 2022 Experimental {B}ayesian
  estimation of quantum state preparation, measurement, and gate errors in
  multiqubit devices {\em Phys. Rev. Res.\/} {\bf 4} 013199

\bibitem{qiskit2024}
Javadi-Abhari A, Treinish M, Krsulich K {\em et~al\/} 2024 Quantum computing
  with {Q}iskit {\em arXiv:2405.08810 [quant-ph]\/}

\bibitem{Zoltan}
Maciejewski F~B, Zimbor{\'{a}}s Z and Oszmaniec M 2020 Mitigation of readout
  noise in near-term quantum devices by classical post-processing based on
  detector tomography {\em {Quantum}\/} {\bf 4} 257

\bibitem{Weisberg2014}
Weisberg S 2014 {\em Applied Linear Regression\/} 4th ed (John Wiley \& Sons,
  Inc.) {A}ppendix A.11, p. 309

\bibitem{Nielsen2021gatesettomography}
Nielsen E, Gamble J~K, Rudinger K {\em et~al\/} 2021 Gate {S}et {T}omography
  {\em Quantum\/} {\bf 5} 557

\bibitem{SmolinEfficient}
Smolin J~A, Gambetta J~M and Smith G 2012 Efficient method for computing the
  maximum-likelihood quantum state from measurements with additive {G}aussian
  noise {\em Phys. Rev. Lett.\/} {\bf 108} 070502

\bibitem{BoEfficient}
Yang B, Raymond R and Uno S 2022 Efficient quantum readout-error mitigation for
  sparse measurement outcomes of near-term quantum devices {\em Phys. Rev. A\/}
  {\bf 106} 012423

\bibitem{PhysRevA.106.012439}
Laflamme R, Lin J and Mor T 2022 Algorithmic cooling for resolving state
  preparation and measurement errors in quantum computing {\em Phys. Rev. A\/}
  {\bf 106} 012439

\bibitem{pereira_parallel_2023}
Pereira L, García-Ripoll J~J and Ramos T 2023 Parallel tomography of quantum
  non-demolition measurements in multi-qubit devices {\em npj Quantum Inf.\/}
  {\bf 9} 22

\bibitem{Geller_2021}
Geller M~R and Sun M 2021 Toward efficient correction of multiqubit measurement
  errors: pair correlation method {\em Quantum Sci. Tech.\/} {\bf 6} 025009

\bibitem{arrasmith2023developmentdemonstrationefficientreadout}
Arrasmith A, Patterson A, Boughton A {\em et~al\/} 2024 Development and
  demonstration of an efficient readout error mitigation technique for use in
  {NISQ} algorithms {\em Proc.\ IEEE International Conference on Quantum
  Computing and Engineering (QCE)\/} p 1244

\bibitem{PRXQuantum.6.010307}
Hashim A, Carignan-Dugas A, Chen L {\em et~al\/} 2025 Quasiprobabilistic
  readout correction of midcircuit measurements for adaptive feedback via
  measurement randomized compiling {\em PRX Quantum\/} {\bf 6} 010307

\end{thebibliography}

\end{document}